\documentclass[12pt]{article}

\usepackage[utf8]{inputenc}
\usepackage{amsmath}
\usepackage{amsthm}   
\usepackage{amssymb}
\usepackage{graphicx}
\usepackage[colorinlistoftodos]{todonotes}
\usepackage{caption}
\usepackage{subcaption}
\usepackage{enumerate}
\usepackage{lscape}
\usepackage{verbatim}
\usepackage{csvsimple}
\usepackage{multirow}
\usepackage{textcomp}
\usepackage{url}
\usepackage{natbib}
\usepackage[linesnumbered,ruled,vlined]{algorithm2e}
\usepackage{bm}
\usepackage{wrapfig}
\usepackage{pdfpages, setspace}
\usepackage{booktabs}
\usepackage{adjustbox}

\newcommand{\wt}{\widetilde}
\newcommand{\wh}{\widehat}

\newcommand{\bQ}{Q}

\newcommand{\bS}{S}
\newcommand{\bX}{X}

\newcommand{\bZ}{Z}
\newcommand{\bbeta}{\beta}

\newcommand{\bbP}{{\mathbb{P}}}
\newcommand{\bbG}{{\mathbb{G}}}

\newcommand{\bOmega}{\Omega}
\newcommand{\bPsi}{\Psi}
\newcommand{\bGamma}{\Gamma}
\newcommand{\bxi}{\xi}
\newcommand{\bSigma}{\Sigma}

\newtheorem{theorem}{Theorem}

\newtheorem{Proposition}{Proposition}

\newcommand{\blind}{1}

\begin{document}

\def\spacingset#1{\renewcommand{\baselinestretch}%
{#1}\small\normalsize} \spacingset{1}

\if1\blind
{
  \title{\bf Cumulative Marginal Mean Model for Assessing Sequential Effects Using Digital Health Data}
  \author{Xingche Guo\\
    Department of Statistics, University of Connecticut\\
    Zexi Cai\\
    Department of Biostatistics, Columbia University\\
    Yuanjia Wang\\
     Departments of Biostatistics and Psychiatry, Columbia University\\
      Donglin Zeng \\
    Department of Biostatistics, University of Michigan \\}
     \date{}
  \maketitle
} \fi

\if0\blind
{
 \title{\bf Cumulative Marginal Mean Model for Assessing Sequential Effects Using Digital Health Data}
      \date{}
  \maketitle
} \fi

\bigskip
\begin{abstract}

Mobile health (mHealth) leverages digital technologies, such as mobile phones, to capture objective, frequent, and real-world digital phenotypes from individuals,
enabling the delivery of tailored interventions to accommodate substantial between-subject and temporal heterogeneity. However, evaluating heterogeneous treatment effects (HTEs) using digital phenotype data is challenging because treatments are delivered dynamically over time and may generate carryover effects that persist beyond the immediate response. Additionally, modeling observational data is complicated by confounding factors. To address these challenges, we propose a double machine learning (DML) method for estimating time-varying HTEs using digital phenotypes under a cumulative marginal mean model that separates current instantaneous effects from lagged carryover effects.
Our approach uses a sequential estimation procedure together with Neyman-orthogonal scores to obtain robust inference for the time-varying HTEs.
We establish the asymptotic normality of the proposed estimator.
Extensive simulation studies validate the finite-sample performance of our approach, demonstrating the advantages of DML and the decomposition of treatment effects.
We apply the method to an mHealth study of Parkinson's disease (PD), where we find that treatment is significantly more effective for younger patients. Our results highlight the potential of the proposed approach for advancing precision medicine in mHealth studies.

\end{abstract}

\noindent%
\vfill

\newpage
\spacingset{1.9} 

\section{Introduction}
\vspace{-0.75em}

The rapid advancement of mobile health (mHealth) technologies holds promise to revolutionize healthcare through the development of digital phenotypes, which are health-related data collected from interactions with digital technologies, including smartphones and wearable sensors. Extending beyond traditional clinical measures, digital phenotypes can provide comprehensive, real-time information on the manifestations and management of illness \citep{jain2015digital}. 
This innovation is relevant in addressing the challenges posed by chronic neurological disorders, such as Parkinson’s disease (PD), Alzheimer’s disease (AD), and dystonia, where access to adequate neurological care remains a major challenge \citep{willis2011neurologist}.
One example of using digital phenotypes to study neurological disorders is the Mobile Parkinson’s Observatory for Worldwide Evidence-based Research \citep[mPower,][]{bot2016mpower}. In this study, researchers remotely collected digital information on daily symptom assessments (e.g., movement, cognition, and voice) from PD patients via smartphones. Another mHealth example is to use wearable devices in cardiac care by continuously monitoring heart rhythms to detect arrhythmia or predict heart failure \citep{alugubelli2022wearable,bhaltadak2024comprehensive}. 
Since digital phenotypes enable continuous, real-time health monitoring and assessments, mobile interventions have been developed in a range of applications. For example,  mobile platforms are designed for diabetes self-management \citep{wu2017mobile}, including text messaging and telemedicine to support patient care.  Behavioral health interventions, such as self-guided cognitive behavioral therapy (CBT) delivered via mobile phones, have also been shown to effectively treat anxiety and depression \citep{martinengo2021self}.

One key advantage of mHealth interventions is their flexibility and adaptivity to different individuals and adaptation over time for the same individual. This feature is important given the considerable heterogeneity among patients and the time-varying nature of heterogeneous treatment effects (HTEs). Patient characteristics, including age and comorbidities, as well as time-specific factors such as real-time health states (e.g., symptom fluctuations, stress levels), can significantly influence responses to interventions. In addition, variability in adherence, engagement, and contextual factors, including environment, social interactions, and timing of delivery, further contribute to heterogeneity. For instance, a behavioral intervention may be more effective when delivered during periods of stability than high stress. In the motivating mPower study, patients received long-term dopaminomimetic therapy (e.g., levodopa) to manage their daily motor symptoms. It is important to estimate the dynamic effects of interventions using digital phenotype data in mHealth studies to improve care. For example, by capturing these dynamics,  adaptive intervention strategies can be developed to tailor treatments to individual patients, improving efficacy and patient engagement. Furthermore, exploring the temporal patterns in treatment effects also facilitates better intervention delivery based on patient's needs, thus advancing personalized medicine and the success of mHealth interventions.

For many mHealth applications, dynamic HTE estimation must account for carryover effects, whereby treatment administered at one time point continues to affect outcomes at subsequent times.
In PD, levodopa rapidly improves motor symptoms by replenishing dopaminergic signaling. However, its therapeutic benefit is typically short-lived and may wane before the next scheduled dose, resulting in end-of-dose ``wearing-off'' and only a short carryover effect into the next dosing interval \citep{pahwa2009levodopa}.
By contrast, interventions in other mHealth settings may induce longer-lasting carryover effects. For instance, a physical activity or exercise session delivered through a mobile app may continue to affect later outcomes such as sleep, fatigue, pain, or mood, rather than exerting its effect only at the time of completion \citep{kubitz1996effects}. In these settings, the outcome observed at a given time point may reflect both the immediate effect of the current treatment and the residual effects of earlier treatments. These examples underscore the need for methods that explicitly accommodate carryover effects in dynamic HTE estimation.

Methods have been developed to estimate treatment effects using longitudinal or time-series data. In particular, the difference-in-differences (DiD) method is a widely used tool to draw causal inferences under the parallel trends assumption, i.e., in the absence of treatment, treated and untreated groups would have similar trends; see the review in \cite{callaway2021difference}. Various extensions have been proposed to relax this assumption by weighting different subjects and time periods, including synthetic control \citep{abadie2010synthetic} and its variations \citep[e.g.,][]{arkhangelsky2021synthetic}. However, these approaches do not handle carryover treatment effects. Towards this end, \cite{sun2021estimating} modeled carryover treatment effects using a parametric dynamic treatment effect model in staggered treatment adoption settings, providing cohort-specific average treatment effects for each cohort and relative time period. \cite{hatamyar2023machine} proposed a Machine Learning Difference-in-Differences (MLDID) method for time-varying conditional average treatment effects on the treated, accounting for dynamic heterogeneity across subpopulations and over time.  \cite{lewis2021double} extended the G-estimation framework for structural nested mean models to dynamic treatment regimes and proposed robust estimation of dynamic effects in high-dimensional settings using double machine learning \citep{chernozhukov2018double}. 
{\cite{loh2023estimating} presented a G-estimation to assess the causal effects of time-varying treatments in longitudinal studies by appropriately adjusting for time-dependent confounders.}
\citet{boruvka2018assessing} introduced the causal excursion effect as a framework for assessing time-varying causal effect moderation in mobile health studies. This framework was later extended by \cite{qian2020linear} and \cite{shi2025meta}.
However, none of these aforementioned methods separate immediate treatment effect from carryover treatment effect due to prior treatment periods.


In this paper, we extend a double machine learning (DML) method \citep{chernozhukov2018double} to estimate time-varying HTEs using digital phenotype data to ensure the robustness of treatment effect estimates to the misspecification of nuisance parameters (e.g., propensity scores and outcome models). DML methods have been used to estimate HTEs for single time-point settings, such as causal forests \citep{wager2018estimation} and R-Learner \citep{nie2021quasi}.  
In this work, we propose a cumulative marginal mean (CUMM) model that expresses the marginal mean potential outcome at each time point as the sum of three components: a baseline mean component, a cumulative direct effects from interventions, and an additional prognostic component.
This decomposition leads to a more accurate estimation of HTEs to capture how the treatment effect manifests over time. We use machine learning (ML) models, such as random forests and gradient boosting machines, to estimate both the baseline and prognostic effects, ensuring flexibility and robustness in capturing complex relationships. The time-varying HTEs are modeled through sequential updates, allowing the method to adapt to changes over time. 
A flexible framework is adopted to capture the carryover treatment effects. 
To address confounding in mHealth studies, we also model the propensity scores using ML models. Finally, we apply the DML framework to ensure unbiased estimation of the treatment effects, leveraging Neyman orthogonalization to control for confounding factors and protect against model misspecification. This methodology enhances the accuracy and reliability of effect estimates in dynamic treatment settings, particularly when utilizing rich digital phenotype data.

The remainder of the article is organized as follows. Section \ref{sec:meth} introduces the proposed CUMM model and presents the framework for estimating time-varying HTEs.
Section \ref{sec:estimation} outlines the DML method for estimation and demonstrates that the proposed estimating equation satisfies the Neyman orthogonality condition. Section \ref{sec:theory} presents the asymptotic theory and the statistical inference procedure. Section \ref{sec:simulation} provides the results of our simulation studies, while Section \ref{sec:application} applies the method to analyze digital phenotype data from an mHealth study. Finally, Section \ref{sec:discussion} summarizes our approach and main findings and discusses potential extensions for future research.


\section{Model}
\vspace{-0.75em}
\label{sec:meth}

We propose a general framework based on the cumulative marginal mean (CUMM) model for assessing sequential effects in digital health data, where treatments are administered at each discrete time $t=1,2,\dots,T$, and $T$ is the maximum number of time points. At each time, each subject has a single outcome measurement, which may be recorded either after the subject receives treatment (e.g., a subject has just completed a session of CBT delivered via a mobile phone or has just taken a levodopa medication) or when no treatment is administered at that time.

Let $Z_i$ denote the baseline covariates for subject $i$, which are assumed to be time-invariant (e.g., race, sex, and age). For subject $i$ at time $t$, let $A_{it}\in\{0,1\}$ denote the treatment indicator, where $A_{it}=1$ indicates that treatment is administered at time $t$, and $A_{it}=0$ otherwise. Let $X_{it}$ denote the pre-specified time-varying effect modifiers of interest. 
We assume that $X_{it}$ is exogenous, meaning that it is observed prior to treatment assignment $A_{it}$ and is not itself affected by $A_{it}$.

We index potential outcomes by treatment history. Let $\bar{A}_{it}=(A_{i1},\dots,A_{it})$ denote the treatment history up to time $t$, and let $Y_{it}(\bar{a}_{it})$ denote the potential outcome for subject $i$ at time $t$ under treatment history $\bar{A}_{it}=\bar{a}_{it}$.
To analyze digital health data, the CUMM model assumes that $E\{Y_{i1}(0)\mid Z_i\} = g_0(Z_i)$; for $t>1$
\vspace{-0.75em}
\begin{align}
E\!\left\{Y_{it}(\bar{a}_{it}) \mid Z_i, \bar X_{it}, {\color{black}\bar{A}_{i,t-1}=\bar{a}_{i,t-1}} \right\}
&=
g_0(Z_i)
+
\sum_{k=1}^t a_{ik}\, \gamma^{t-k}\,\beta_{0k}^\top \widetilde{X}_{ik}
+
\delta_{0t}(\widetilde{X}_{it}), \label{equ:yt0_revised}
\end{align}
where $\widetilde{X}_{it}=(1,Z_i^\top,X_{it}^\top)^\top,$ $\bar{X}_{it} = (X_{i1}, \dots, X_{it})$, $\gamma \in (0, 1]$. It follows from model \eqref{equ:yt0_revised} that the time-varying instantaneous treatment effect at time $t$ can be expressed as
\vspace{-0.75em}
\begin{align}
E\!\left\{Y_{it}(\bar{a}_{i,t-1},1)-Y_{it}(\bar{a}_{i,t-1},0)\mid Z_i,\bar{X}_{it},{\color{black}\bar{A}_{i,t-1}=\bar{a}_{i,t-1}}\right\}
=
\beta_{0t}^{\top}\widetilde{X}_{it},
\label{equ:hte_revised}
\end{align}
From \eqref{equ:hte_revised}, we assume that the effects of past treatments are fully captured through the cumulative carryover contributions of prior treatment episodes $(k < t)$, whereas the instantaneous treatment effect at time $t$ depends only on the baseline covariates and the current effect modifiers.

The first term on the right-hand side of \eqref{equ:yt0_revised} denotes the baseline mean outcome under no treatment, conditional on the baseline covariates.
The second term represents the cumulative direct effects from interventions that captures both the instantaneous effect of the current treatment and the carryover effects of prior treatments. As $\gamma \to 0$, the carryover treatment effect disappears, implying that the treatment effect is purely short-term and fully dissipates before the next treatment occasion \citep{xu2023mixed}. At the other extreme, when $\gamma = 1$, the treatment effect accumulates over time without any decay. In our applications, we assume an exponential decay pattern, though the proposed framework can be generalized to allow for alternative decay rates.
The third term, $\delta_{0t}(\wt X_{it})$, is an unspecified prognostic function that captures additional time-varying changes in the outcome process not explained by the baseline covariates or the accumulated treatment effects. Moreover, $\delta_{0t}(\wt X_{it})$ is treated as a nuisance function and does not need to be estimated.
Thus, model \eqref{equ:yt0_revised} assumes that the marginal mean of the potential outcome at time $t$ admits an additive decomposition into the baseline mean, the cumulative carryover effects of current and previous treatments, and the current prognostic state.

Let \(U_{it}\) denote a low-dimensional vector of time-varying concurrent variables observed prior to treatment assignment at time \(t\) (e.g., symptom severity and wearable-based physiological measurements). We assume that \((Z_i,X_{it})\) is a subvector of \(U_{it}\). 
To identify and estimate the coefficients \(\beta_{0t}\) from observational data, we impose the following assumptions:

\noindent\textsc{Condition 1}. For each subject \(i\) and time point \(t\), assume that:
\begin{enumerate}

\item[(a)] \textit{Consistency:}
$
Y_{it}
=
\sum_{\bar a_t\in\{0,1\}^t}
I(A_{i1}=a_1,\ldots,A_{it}=a_t)\,Y_{it}(\bar a_t).
$

\item[(b)] \textit{Sequential ignorability:} Let $H_{it}$ consist of all historical information including $(\bar U_{it},\, \bar A_{i,t-1},\, \bar X_{it})$. Then 
$
    A_{i t} \perp \left\{Y_{i t}\left(\bar{A}_{i, t-1}, a_t\right): a_t \in\{0,1\}\right\} \mid H_{i t} .
$

\item[(c)] \textit{No additional effect modification:}
\begin{align*}
&E\!\left[
Y_{it}(\bar a_{i,t-1},1)-Y_{it}(\bar a_{i,t-1},0)
\mid \bar A_{i,t-1}=\bar a_{i,t-1},\bar X_{i, t-1},U_{it}
\right] \\
&\qquad =
E\!\left[
Y_{it}(\bar a_{i,t-1},1)-Y_{it}(\bar a_{i,t-1},0)
\mid Z_i,\bar A_{i,t-1}=\bar a_{i,t-1},\bar X_{it}
\right].
\end{align*}

\item[(d)] \textit{Concurrent dependence:} 
    $P(A_{it}\mid H_{it})=P(A_{it}\mid U_{it})$.
\end{enumerate}

Condition 1(a) is the standard consistency assumption, and Condition 1(b) is a sequential ignorability assumption stating that, conditional on the observed history \(H_{it}\), treatment assignment at time \(t\) is independent of all future potential outcomes. 
Condition 1(c) states that among the components of \(U_{it}\), only $(Z_i, X_{it})$ acts as an effect modifier, whereas the remaining components are prognostic in the sense that they may predict the outcome level but do not alter the treatment contrast. The partially linear model in the double/debiased machine learning approach in \cite{chernozhukov2018double} also satisfies this condition.
Condition 1(c) is often plausible in mHealth applications. In such settings, \(U_{it}\) may include time-varying patient status variables, such as recent symptoms, behavioral patterns, or physiological states, 
that might be affected by prior treatments and predictive of the current outcome. In many settings, these variables primarily reflect the subject's current pre-treatment state at time $t$, rather than modifying the effect of the newly delivered intervention.
Condition 1(d) formalizes the idea that the treatment assignment mechanism depends on the concurrent history only through \(U_{it}\). This assumption is especially natural in settings where interventions are administered in real time based primarily on the subject's current status.
Condition 1 is central to the identification strategy, justify the proposed estimation procedure, and are used to establish the Neyman orthogonality of the score in Section~\ref{sec:estimation}.

\vspace{-1em}
\section{Estimation and Inference}
\vspace{-1em}
\label{sec:estimation}

We now describe a sequential double machine learning (DML) procedure for estimating $\beta_{t}$ under the CUMM model. Because the untreated potential outcome at time $t$ depends on the accumulated carryover effects of treatments delivered at earlier time points, the estimation must be carried out recursively over time. The main idea is to combine flexible machine-learning methods for estimating nuisance functions with Neyman-orthogonal score equations to obtain robust inference for the low-dimensional target parameter $\beta_{0t}$.

We first estimate the baseline mean function $g_0(Z_i)=E\{Y_{i1}(0)\mid Z_i\}$.
Under consistency, for individuals with $A_{i1}=0$, we observe $Y_{i1}=Y_{i1}(0)$. Therefore, $g_0$ can be estimated by minimizing the empirical squared loss
$n^{-1}\sum_{i=1}^n
I(A_{i1}=0)\bigl\{Y_{i1}-g(Z_i)\bigr\}^2$,
where $g(\cdot)$ may be estimated using flexible nonparametric or machine-learning methods, such as splines, random forests, or boosting. Let $\widehat g$ denote the resulting estimator.

Next, we estimate the propensity score $\pi_{0t}(U_{it})
 = P(A_{it}=1\mid U_{it})$
using flexible machine-learning methods, such as gradient boosting or neural networks, trained on observations pooled across all decision times $t$, i.e., $\pi_{0t}(U_{it}) = \pi_{0}(U_{it}, t)$.
This is a working model used to improve efficiency and stabilize estimation when 
$T$ is moderate or large.
Let $\widehat\pi_t(U_{it})$ denote the estimated propensity score at time $t$.

Define the residualized outcome by removing the baseline component and the cumulative treatment contributions up to time $t$:
\vspace{-0.8em}
\begin{align*}
    R_{it} = R_{it}(\bar A_t), \qquad R_{it}(\bar a_t)
=
Y_{it}(\bar a_t)
-
g_0(Z_i)
-
\sum_{k=1}^{t} a_k\gamma^{t-k}\beta_{0k}^{\top}\widetilde X_{ik}.
\end{align*}
We then formalize the nuisance prognostic function through the following proposition.
\vspace{-0.8em}
\begin{Proposition}\label{prop:residual}
Under Condition~1, $E(R_{it}\mid U_{it},A_{it})
=
E(R_{it}\mid U_{it}).$
\end{Proposition}
\vspace{-0.8em}
Proposition~\ref{prop:residual} shows that, conditional on \(U_{it}\), the residualized outcome has the same mean under both treatment levels. The proof is given in the Supplementary Material. This motivates defining the nuisance prognostic function as
$\mu_{0t}(U_{it}) = E(R_{it}\mid U_{it}).$

To estimate the instantaneous treatment effect at time $t$, note that the untreated mean function depends on the carryover contributions from earlier treatment effects. We therefore proceed sequentially in $t$. For now, we assume that the discount parameter $\gamma$ is known; selection of $\gamma$ is discussed in Remark 1 below.
Suppose that, for some $t>1$, we have already obtained estimators $\widehat{\beta}_1,\dots,\widehat{\beta}_{t-1}$, the estimation proceeds in two steps:

\noindent Step 1.
Using observations with \(A_{it}=0\), estimate the nuisance function \(\mu_{0t}(\cdot)\) by regressing the estimated residualized outcome before removing the current treatment effect
\begin{align*}
    \widehat R_{it}^{(0)}
=
Y_{it}
-
\widehat g(Z_i)
-
\sum_{k=1}^{t-1} A_{ik}\gamma^{t-k}\widehat\beta_{k}^{\top}\widetilde X_{ik}
\end{align*}
on \(U_{it}\), that is, by minimizing
\[
\ell_t(\mu_t)
=
\frac{1}{n}\sum_{i=1}^n
I(A_{it}=0)
\Big\{
\widehat R_{it}^{(0)}-\mu_t(U_{it})
\Big\}^2.
\]
Let \(\widehat\mu_{t}\) denote the resulting estimator.

\noindent  Step 2: Estimation of $\beta_{0t}$ via DML.
We estimate $\beta_{0t}$ by solving the moment equation
\begin{align}
\frac{1}{n}\sum_{i=1}^n
\bigl\{A_{it}-\widehat\pi_t(U_{it})\bigr\}
\Bigl[
\widehat R_{it}^{(0)}-\widehat\mu_t(U_{it})-A_{it}\beta_t^\top \widetilde X_{it}
\Bigr] \widetilde X_{it}
=0.
\label{eq:DML_revised}
\end{align}
The solution is denoted by $\widehat{\beta}_t$.

We now provide intuition for why the proposed estimator is robust to regularization bias in the nuisance estimators. The key property is that the score function in \eqref{eq:DML_revised} satisfies Neyman orthogonality.
Let $(Z, A_t,Y_t, U_t, X_t)$ denote a generic copy of the observed data at time $t$. 
Define
\begin{align}
h_t\!\left(H_{t}; g,\mu_t,\{\beta_k\}_{k=1}^{t-1}\right)
=
g(Z)
+
\sum_{k=1}^{t-1}
A_{k}\gamma^{t-k}\,
\beta_k^\top \widetilde X_{k}
+
\mu_t(U_{t}).
\label{eq:ht_def}
\end{align}
Define the score
\vspace{-2em}
\begin{align}
S_t(\beta_t,\eta_t)
=
\Big\{
Y_t-A_t\beta_t^\top \widetilde X_t - h_t(H_t)
\Big\}
\Big\{
A_t-\pi_t(U_t)
\Big\}
\widetilde X_t,
\label{eq:score}
\end{align}
where $\eta_t=(h_t,\pi_t)$.
The corresponding population moment is $\Psi_t(\beta_t,\eta_t)=E\{S_t(\beta_t,\eta_t)\}$,
and its empirical realization is $\widehat\Psi_t(\beta_t,\eta_t)=\bbP_n S_t(\beta_t,\eta_t)$, where $\bbP_n$ is the empirical measure.

\vspace{-0.8em}
\begin{Proposition}\label{prop:neyman}
Under Condition 1:
\begin{enumerate}
\item[(i)] The score is unbiased at the truth:
$\Psi_t(\beta_{0t},\eta_{0t})=E\{S_t(\beta_{0t},\eta_{0t})\}=0.$

\item[(ii)] The score is Neyman orthogonal with respect to the nuisance parameter \(\eta_t\) at \((\beta_{0t},\eta_{0t})\). Specifically, for any perturbation direction $\Delta\eta_t=(\Delta h_t,\Delta\pi_t)$,
the Gateaux derivative satisfies
$\left.
\frac{\partial}{\partial \rho}
\Psi_t\bigl(\beta_{0t},\eta_{0t}+\rho\,\Delta\eta_t\bigr)
\right|_{\rho=0}
=0.$
\end{enumerate}
\end{Proposition}

Proposition~\ref{prop:neyman} establishes the Neyman orthogonality property, and its proof is provided in the Supplementary Materials. 
It implies that the score is locally insensitive, to first order, to estimation errors in the nuisance functions $h_{0t}$ and $\pi_{0t}$. Consequently, the estimator $\widehat{\bbeta}_t$ can remain consistent and asymptotically normal even when the nuisance estimators are obtained using flexible machine-learning procedures with slower convergence rates; see Section \ref{sec:theory} for the formal results.

\noindent\textbf{Remark 1}. When the carryover discounting factor $\gamma$ is unknown, we treat it as a tuning parameter. Specifically, the root mean squared error (RMSE) between the true and estimated responses is computed for a given discounting factor $\gamma$, denoted as $\mathcal{R}(\gamma)$, and the optimal $\gamma$ is determined by minimizing $\mathcal{R}(\gamma)$. Alternatively, other criteria, such as the root mean squared percentage error (RMSPE) or cross-validation (CV) scores, can also be used to select $\gamma$.

\noindent\textbf{Remark 2.}
The sequential estimator may be statistically inefficient because, at each time point \(t\), the nuisance function \(\mu_{0t}\) is estimated using only the observations collected at that same time point. To improve efficiency, one may impose an additional smoothness assumption that \(\mu_{0t}\) evolves gradually over time. Under this assumption, observations from time points near \(t\) can be borrowed to estimate \(\mu_{0t}\) more efficiently while still preserving the sequential nature of the procedure. A more detailed discussion can be found in the Supplementary Materials.


\vspace{-1em}
\section{Asymptotic Theory}
\label{sec:theory}

Following \citet{vaart1996weak}, let \(P\) denote the true data-generating distribution, with corresponding population and empirical measures \(\mathbb{P}\) and \(\mathbb{P}_n\). For a generic observation \(O\sim P\) and observed data \(O_1,\dots,O_n\),
$
\mathbb{P}f=E\{f(O)\}=\int f\,dP$ and $\mathbb{P}_n f=\frac{1}{n}\sum_{i=1}^n f(O_i).$
Define
\vspace{-0.8em}
\[
\Psi_t^{a}(\eta_t)
=
\mathbb{P}_n\!\left[
A_t\{A_t-\pi_t(U_t)\}\widetilde X_t\widetilde X_t^\top
\right], \qquad
\Psi_t^{b}(\eta_t)
=
\mathbb{P}_n\!\left[
\{Y_t-h_t(H_t)\}\{A_t-\pi_t(U_t)\}\widetilde X_t
\right].
\]
Define $\widehat{ {J}}_{t} = \Psi_t^{a}( \widehat{  {\eta}}_{t})$. According to \eqref{eq:DML_revised}, $\wh{\bbeta}_t$ has an explicit form: $\wh{\bbeta}_t = \widehat{ {J}}_{t}^{-1} \Psi_t^{b}( \wh{  {\eta}}_t)$. 
Denote ${\cal T} = \{1, 2, \dots, T\}$, assume that $\cal T$ is a finite set. Assume that the dimension of $\wt{\bX}_t$, $d$, is finite.
We establish the asymptotic conditions under which $n^{1/2} ( \widehat{\bbeta}_t - \bbeta_{0t} )$ converges in distribution to a Gaussian limit uniformly for all $t \in {\cal T}$. The following conditions are needed for the theorems in this paper. 

\noindent\textsc{Condition 2}. For $t \in {\cal T}$,  $p_a < \pi_{0t}(U_{t}) < 1-p_a$ for small $ p_a > 0$, and $E \left( R_t^2 \mid U_{t} \right) < \infty$. 

\noindent\textsc{Condition 3}. Define $\Gamma_t = E [ \widetilde X_{t} \widetilde X_{t}^{\top} ]$, let $\Lambda_{\min}(\cdot)$ and $\Lambda_{\max}(\cdot)$ denote the minimum and maximum eigenvalues for a matrix, then, $0 < \sigma_{\min}^2 \le \inf_{t \in {\cal T}} \Lambda_{\min}\left( \Gamma_t \right)  \le \sup_{t \in {\cal T}} \Lambda_{\max}\left( \Gamma_t \right) \le \sigma_{\max}^2 < \infty$.
Furthermore, define $\Upsilon_t = E [ ( \widetilde X_{t} \widetilde X_{t}^{\top}) \otimes (\widetilde X_{t} \widetilde X_{t}^{\top}) ]$, we have $\sup_{t \in {\cal T}} \Lambda_{\max}\left( \Upsilon_t \right) \le \kappa_{\max}^4 < \infty$.

\noindent\textsc{Condition 4}. 
Suppose $\wh{\pi}_t \in \mathcal{F}_{\pi}$, $\wh{g} \in \mathcal{F}_{g}$, $\wh{\mu}_t \in \mathcal{F}_{\mu}$, where $\mathcal{F}_{\pi}$, $\mathcal{F}_{g}$, $\mathcal{F}_{\mu}$ are $P$-Donsker. Furthermore
\vspace{-1em}
$$\bbP | \wh{g} - g_{0} |^4 = o_p(1), \quad  \sup_{t \in \cal T} \bbP | \wh{\mu}_t - \mu_{0t} |^4 = o_p(1), \quad
\sup_{t \in \cal T} \bbP | \wh\pi_t - \pi_{0t} |^4 = o_p(1),$$
$$
\sup_{t \in \cal T} \bbP | (\wh{\mu}_t - \mu_{0t}) ( \wh{\pi}_t - \pi_{0t}) |^2 = o_p(n^{-1}),\ \quad \sup_{t \in \cal T} \bbP | (\wh{g} - g_0) ( \wh{\pi}_t - \pi_{0t}) |^2 = o_p(n^{-1}).
$$

 Condition 2 ensures the effective sample size at any given time point $t$, both treated and untreated, is of the same order as the total number of subjects $n$.
 Condition 3 imposes uniform nondegeneracy and bounded moment conditions on $\wt X_t$. Condition 3 is mild, as it is readily satisfied by Gaussian random variables.
 Condition 4 imposes constraints on the model complexity of the machine learning estimators, requires the consistency of the estimates, and specifies the order of the estimation errors.
    Moreover, by the definition of $h_t(\cdot)$, imposing assumptions on $g(\cdot)$ and $\mu_t(\cdot)$ is equivalent to imposing assumptions on $h_t(\cdot)$. Specifically, for $\wh{h}_t \in \mathcal{F}_{h_t}$, $\mathcal{F}_{h_t}$ is $P$-Donsker.  Also, $\sup_{t \in \cal T} \bbP | \wh{h}_t - h_{0t} |^4 = o_p(1)$, $\sup_{t \in \cal T} \bbP | (\wh{h}_t - h_{0t}) ( \wh{\pi}_t - \pi_{0t}) |^2 = o_p(n^{-1})$.


\begin{theorem} \label{thm:main1}
Under Conditions 1–4, as $n \rightarrow \infty$, the estimator $\wh\bbeta = (\wh\bbeta_1^{\top}, \dots, \wh\bbeta_T^{\top})^{\top}$ has the following asymptotic expression
\begin{align*}
     n^{1/2} \left( \widehat \bbeta - \bbeta_{0} \right) \overset{d}{\rightarrow} N \left(  {0}, \mathrm{J}_0^{-1} \bPsi_{0} \mathrm{J}_0^{-1} \right),
\end{align*}
where  $\bbeta_0 = (\bbeta_{01}^{\top}, \dots, \bbeta_{0T}^{\top})^{\top}$,
$\mathrm{J}_0 = \mathrm{diag}( {J}_{0t})_{t \in{\cal T}}$, $\bPsi_0 = (\bOmega_{0,tt'})_{t,t' \in {\cal T}}$,
${J}_{0t} = E\left[  A_t \left\{A_{t}-\pi_{0t}(U_{t}) \right\} \widetilde X_{t} \widetilde X_{t}^{\top} \right]$, $\bOmega_{0,tt'} =  E\big[ S_{t}(\bbeta_{0t},   {\eta}_{0t}) S_{t'}^{\top}(\bbeta_{0t'},   {\eta}_{0t'}) \big]$. 
\end{theorem}

The proof of the theorem is in the supplementary materials. By the definition of ${J}_{0t}$, we have
${J}_{0t} =E\left[  \pi_{0t}(U_{t}) (1 - \pi_{0t}(U_{t})) \widetilde X_{t} \widetilde X_{t}^{\top} \right]$.
Under Conditions 2 and 3, $ {J}_{0t}$ is positive definite with eigenvalues bounded away from zero and infinity. Meanwhile, 
\begin{align*}
    \bOmega_{0,tt} = E\left[\left\{R_{t}-\mu_{0t}(U_{t}) \right\}^2 \left\{A_{t}-\pi_{0t}(U_{t}) \right\}^2 \widetilde X_{t} \widetilde X_{t}^{\top} \right],
\end{align*} 
and it can similarly be shown that $\bOmega_{0,tt}$ is positive definite with bounded eigenvalues. This ensures that the covariance of the limiting Gaussian distribution is finite, allowing us to conduct statistical inference as follows:
Define $\wh{\mathrm{J}} = \mathrm{diag}(\wh{ {J}}_{t})_{t \in {\cal T}}$, 
$\widehat{ {J}}_{t} = \Psi_t^{a}( \widehat{  {\eta}}_{t})$,
$\wh{\bPsi} = (\wh{\bOmega}_{tt'})_{t,t' \in {\cal T}}$,
$\wh{\bOmega}_{tt'} =  \bbP_n \big[ \bS_t(\wh{\bbeta}_{t}, \wh{  {\eta}}_{t}) \bS_{t'}^{\top}(\wh{\bbeta}_{t'}, \wh{  {\eta}}_{t'}) \big]$. We can consistently estimate the variance-covariance matrix for  $n^{1/2} \left( \widehat \bbeta - \bbeta_{0} \right)$ as $\widehat{  {\Sigma}}=\wh{\mathrm{J}}^{-1} \wh\bPsi \wh{\mathrm{J}}^{-1}$. 
Thus, we construct the $(1-\alpha) \times 100\%$ confidence intervals for the $k$-th entry of $\bbeta_{0t}$, denoted as $\bbeta_{0t}^{(k)}$, by: 
\begin{align}
    \wh{\bbeta}_t^{(k)}  \pm z_{1-\alpha/2} \sqrt{ n^{-1} \left( \widehat{ {J}}_{t}^{-1} \wh{\bOmega}_{tt} \widehat{ {J}}_{t}^{-1} \right)_{(k,k)}  }, \label{equ:CI}
\end{align}
where $z_{1 - \alpha/2}$ denotes the $1 - \alpha/2$ quantile of the standard normal distribution.
In addition, to test if there are overall excursion effects (i.e., $\bbeta_{0t} =  {0}$ for all $t \in {\cal T}$), one can construct an overall test based on a Wald-type $\chi^2$-test statistic, $n \wh\bbeta^{\top} \wh{\mathrm{J}} \wh\bPsi^{-1} \wh{\mathrm{J}} \wh\bbeta $, which is approximately $\chi_{Td}^2$ distributed for large $n$.

\noindent\textbf{Remark 3}. 
Our proposed framework yields an estimated sequence of treatment effects, denoted by $\{\widehat{\bbeta}_t\}_{t=1}^T$. In some applications, one may further impose a lower-dimensional parametric model on the treatment-effect trajectory, for example by assuming $\bbeta_t=f(t;\theta)$,
where $f(\cdot;\theta)$ may be specified as a linear, quadratic, or other smooth function of time. Given the estimators $\{\widehat{\bbeta}_t\}_{t=1}^T$ and their estimated covariance matrix $\widehat{\Sigma}$, one may estimate $\theta$ by generalized least squares, namely by minimizing $\bigl\{\widehat{\bbeta}-f(\theta)\bigr\}^{\top}\widehat{\Sigma}^{-1}\bigl\{\widehat{\bbeta}-f(\theta)\bigr\}$
where $f(\theta)
=
\bigl(f(1;\theta)^\top,\dots,f(T;\theta)^\top\bigr)^\top.$
The resulting estimator of $\theta$ is asymptotically normal under standard regularity conditions, by the asymptotic properties of $\widehat{\bbeta}$ established in this section.

Finally, we consider the asymptotic regime in which the number of time points $T$ is large, which is common in mHealth applications. To this end, suppose that the true treatment effect sequence is generated by a smooth function $\bxi_0:[0,1]\to\mathbb{R}^d$ such that $\bbeta_{0t}=\bxi_0(t/T)$ ($t=1,\dots,T$).
Correspondingly, define the piecewise-constant stochastic process $\widehat\bxi(\cdot)$ on $[0,1]$ by $
\widehat\bxi(s)=\widehat\bbeta_t$ for any $s\in((t-1)/T,t/T]$.
Similarly, define the piecewise-constant stochastic process $\widetilde\bGamma(\cdot)$ by $\widetilde\bGamma(s)=J_{0t}^{-1}\bS_t(\bbeta_{0t},\eta_{0t})$ for $s\in((t-1)/T,t/T]$.
In addition, assume that $A_t$, $Y_t$, $U_t$, and $\widetilde\bX_t$ are realizations of stochastic processes indexed by $t/T\in[0,1]$, and that these processes have bounded total variation almost surely. The following theorem establishes the asymptotic behavior of $\widehat\bxi(\cdot)$.

\begin{theorem} \label{thm:main2}
In addition to Conditions 1–4, we assume that $\bxi_0$ is continuously differentiable in $[0,1]$ and that there exists a continuous bivariate function, $\widetilde\bSigma_0$, such that $Cov(\widetilde\bGamma(s), \widetilde\bGamma(s'))$ converges to $\widetilde\bSigma_0(s,s')$ for any $s, s'\in [0,1]$. Assume $\sqrt n/ T\rightarrow 0$. Then $\sqrt n (\wh\bxi(s)-\bxi_0(s))$ weakly converges to a tight Gaussian process with mean zero and covariance function $\widetilde\bSigma_0(s,s')$ in $l^{\infty}[0,1]$.
\end{theorem}

This theorem gives the uniformly weak convergence for the piecewise-constant process given by $\wh\bbeta_t$ when the number of time points is large. One useful application based on this theorem is that we can construct a confidence band or perform global testing over all the time points.

\section{Simulations}
\label{sec:simulation}

We conducted extensive simulation studies to evaluate the finite-sample performance of the proposed method. Baseline covariates $\bZ_i = (Z_{i1}, Z_{i2})$ were generated with each sampled independently from a standard normal distribution. For the time-varying exogenous covariates, $X_{it} = (X_{i1t},  X_{i2t})^{\top}$, we used the following generative model to simulate fluctuating trajectories to mimic subjects' daily physical activities: for $q=1,2$
\begin{align}
    X_{iqt} = \sqrt{2} \sum_{k = 1}^{K} {\xi}_{iqk} \sqrt{\nu_{k}} \cos (k \pi t / T) + \epsilon_{iqt}, \label{equ:gener_X}
\end{align}
where $\xi_{i q k} \sim \text{ i.i.d. } N(0, 1)$ represents the functional principal scores, and $\epsilon_{iqt} \sim \text{ i.i.d. } N(0, \sigma_{\epsilon}^2)$ is the noise term. In our simulations, we set $\sigma_{\epsilon} = 0.75$, $\nu_k = 0.3 \exp(-k/8)$, and $K=30$.
Assume all subjects have the same trial length $T=100$, and let $\rho = t / T$ represent the scaled time. 
Denote $\widetilde X_{it}
= (1,\, \bZ_{i1},\, \bZ_{i2},\, X_{i1t},\, X_{i2t})^\top$.
Define the corresponding instantaneous treatment effects 
as $\bbeta_{0t} = (-0.3 (1 - \rho / 2),\, 0.05,\, -0.05,\, 0.05 + 0.1 \rho^2,\, -0.05 - 0.1 \rho^2)^{\top}$.
Suppose there are additional prognostic variables
$\widetilde U_{it}=(\widetilde U_{i1t},\dots,\widetilde U_{iRt})^\top$,
where each component \(\widetilde U_{irt}\) is generated independently according to \eqref{equ:gener_X}. Define $U_{it}=(\widetilde X_{it},\,\widetilde U_{it})$.
The potential outcome is then generated as
\vspace{-2em}
\begin{align*}
Y_{it}(\bar a_{it})
=
g_0(Z_{i1},Z_{i2})
+
\sum_{k=1}^t \gamma^{t-k} a_{ik}\beta_{0k}^\top \widetilde X_{ik}
+
\mu_{0t}(U_{it})
+
W_{it},
\qquad
W_{it}\overset{\mathrm{i.i.d.}}{\sim}N(0,0.3^2).
\end{align*}
We considered the carryover discounting factors $\gamma \in \{0.3, 0.7\}$. 
To mimic the complexities encountered in real-world studies, our simulation framework includes a missing data component. Specifically, each observation is assumed to be missing with probability $0.3$. The proposed approach remains valid under the missing at random assumption.

We considered two data-generating scenarios:

\textbf{Case I}: In this scenario, we assumed the presence of numerous irrelevant prognostic factors, setting the number of prognostic factors to be $R=8$. 
We defined $\nu_{it} = X_{i1t} + \wt U_{i1t} + \wt U_{i2t}$, which influenced both the propensity score and the prognostic function, by assuming the probability of treatment, i.e., $A_{it}=1$, to be given by $\pi_{0t}( U_{it}) = \pi_{0}( U_{it})= 1 / (1 + \exp( - \frac{1}{2}\nu_{it}))$, and the prognostic function to be $\mu_{0t}(U_{it}) = \mu_{0}(U_{it}) = 0.05 \nu_{it}$. The baseline outcome was specified as $g_0(Z_{i1}, Z_{i2}) = -1.5 + 0.1 (Z_{i1} +Z_{i2})$.

\textbf{Case II}: In this scenario, we considered a more nonlinear relationship for both the propensity score and prognostic function. Here, we set the number of prognostic factors to be $R=2$, $\nu_{it} = 2 \cdot I \left(  \{X_{i1t} > 1\} \cup  \{\wt U_{i1t}>0.2 \}\right) - 1$.
The probability of $A_{it}=1$ was then given by $\pi_{0t}( U_{it}) = \pi_{0}( U_{it}) = 1 / (1 + \exp( \nu_{it}))$, and the prognostic function was defined as
$\mu_{0t}(U_{it}) = \mu_{0}(U_{it}) = 0.15 \nu_{it}$.
The baseline expected outcome was specified as $g_0(Z_{i1}, Z_{i2}) = -1.5 + 0.5 I(|Z_{i1}| > 0.5)$.

\begin{table}[h!]
\caption{Summary of the estimated $\wh\beta_t$ across 500 simulations for Case I.} 
\label{table_simu1}
{\centering
\begin{adjustbox}{max width=\textwidth}
    \begin{tabular}{rlrcccrcccrcccrccc}
    \toprule 
     &      & \multicolumn{8}{c}{$n=200$}                           & \multicolumn{8}{c}{$n=500$} \\
     &      & \multicolumn{4}{c}{$\gamma=0.3$} & \multicolumn{4}{c}{$\gamma=0.7$} & \multicolumn{4}{c}{$\gamma=0.3$} & \multicolumn{4}{c}{$\gamma=0.7$} \\
    \multicolumn{1}{l}{$\wh\beta_t$} & \textbf{Method} & \multicolumn{1}{c}{Bias} & SD   & SE   & CP   & \multicolumn{1}{c}{Bias} & SD   & SE   & CP   & \multicolumn{1}{c}{Bias} & SD   & SE   & CP   & \multicolumn{1}{c}{Bias} & SD   & SE   & CP \\
\cmidrule(lr){3-6}\cmidrule(lr){7-10}\cmidrule(lr){11-14}\cmidrule(lr){15-18}
\multicolumn{1}{l}{Intercept} & Proposed (known) & -0.8 & 5.8  & 5.4  & 92.6 & -0.7 & 5.8  & 5.4  & 92.6 & -0.6 & 3.4  & 3.3  & 93.9 & -0.6 & 3.4  & 3.3  & 94.0 \\
         & Proposed (estimated) & -0.8 & 5.7  & 5.4  & 92.6 & -0.7 & 5.8  & 5.4  & 92.6 & -0.6 & 3.4  & 3.3  & 93.9 & -0.6 & 3.4  & 3.3  & 94.0 \\
         & Direct-DML (2) & -1.4 & 6.0  & 5.2  & 90.3 & 8.1  & 6.6  & 5.6  & 89.7 & -2.2 & 3.5  & 3.3  & 93.3 & 1.0  & 3.8  & 3.5  & 93.2 \\
         & Direct-DML (5) & -4.6 & 6.4  & 5.0  & 87.4 & 10.3 & 6.9  & 5.4  & 86.6 & -3.1 & 3.5  & 3.2  & 92.3 & 2.0  & 3.8  & 3.5  & 92.0 \\
         & No-DML & -4.4 & 5.8  & 3.8  & 79.5 & -4.3 & 5.8  & 3.8  & 79.6 & -4.5 & 3.5  & 2.3  & 79.8 & -4.5 & 3.5  & 2.3  & 79.8 \\[0.6em]
    \multicolumn{1}{l}{$Z_1$} & Proposed (known) & 0.4  & 5.6  & 5.5  & 94.6 & 0.5  & 5.7  & 5.5  & 94.6 & 0.1  & 3.3  & 3.3  & 95.0 & 0.2  & 3.4  & 3.3  & 95.1 \\
         & Proposed (estimated) & 0.5  & 5.6  & 5.4  & 94.6 & 0.3  & 5.7  & 5.5  & 94.6 & 0.2  & 3.3  & 3.3  & 95.1 & 0.1  & 3.4  & 3.3  & 95.0 \\
         & Direct-DML (2) & -5.6 & 5.9  & 5.2  & 89.9 & -9.4 & 6.5  & 5.6  & 89.3 & -1.4 & 3.4  & 3.3  & 93.4 & -3.1 & 3.7  & 3.6  & 93.2 \\
         & Direct-DML (5) & -12.3 & 6.2  & 5.0  & 86.8 & -18.7 & 6.8  & 5.4  & 85.5 & -3.0 & 3.5  & 3.2  & 92.5 & -5.5 & 3.8  & 3.5  & 92.0 \\
         & No-DML & 0.0  & 4.3  & 3.8  & 90.7 & 0.0  & 4.0  & 3.8  & 93.1 & 0.1  & 2.6  & 2.2  & 90.9 & 0.0  & 2.4  & 2.2  & 93.6 \\[0.6em]
    \multicolumn{1}{l}{$Z_2$} & Proposed (known) & 0.3  & 5.6  & 5.5  & 94.4 & 0.2  & 5.7  & 5.5  & 94.4 & -0.1 & 3.4  & 3.3  & 94.9 & -0.1 & 3.4  & 3.3  & 94.7 \\
         & Proposed (estimated) & 0.3  & 5.6  & 5.4  & 94.5 & 0.4  & 5.7  & 5.5  & 94.4 & -0.1 & 3.4  & 3.3  & 94.9 & -0.1 & 3.4  & 3.3  & 94.7 \\
         & Direct-DML (2) & -3.0 & 5.9  & 5.2  & 89.9 & 1.2  & 6.3  & 5.6  & 90.6 & -0.9 & 3.4  & 3.3  & 93.3 & 0.5  & 3.7  & 3.6  & 93.5 \\
         & Direct-DML (5) & -9.3 & 6.1  & 5.0  & 87.0 & -2.0 & 6.4  & 5.4  & 88.8 & -2.2 & 3.5  & 3.2  & 92.4 & -0.1 & 3.8  & 3.5  & 92.7 \\
         & No-DML & -1.5 & 4.4  & 3.8  & 90.3 & -0.9 & 4.0  & 3.8  & 92.8 & 1.7  & 2.6  & 2.2  & 90.4 & 1.0  & 2.4  & 2.2  & 93.3 \\[0.6em]
    \multicolumn{1}{l}{$X_1$} & Proposed (known) & 0.2  & 6.3  & 5.9  & 92.0 & 0.2  & 6.4  & 6.0  & 92.0 & -0.2 & 3.7  & 3.6  & 93.8 & -0.2 & 3.7  & 3.6  & 93.8 \\
         & Proposed (estimated) & 0.2  & 6.3  & 5.9  & 92.0 & 0.2  & 6.4  & 6.0  & 92.0 & -0.2 & 3.7  & 3.6  & 93.9 & -0.2 & 3.7  & 3.6  & 93.8 \\
         & Direct-DML (2) & -3.4 & 6.4  & 5.7  & 90.6 & -4.4 & 7.0  & 6.2  & 90.3 & -1.6 & 3.8  & 3.6  & 93.3 & -2.1 & 4.1  & 3.9  & 93.3 \\
         & Direct-DML (5) & -7.4 & 6.6  & 5.5  & 88.7 & -9.0 & 7.1  & 5.9  & 88.6 & -2.9 & 3.8  & 3.6  & 92.4 & -3.6 & 4.1  & 3.8  & 92.4 \\
         & No-DML & -3.5 & 5.7  & 4.1  & 83.9 & -3.4 & 5.7  & 4.1  & 84.0 & -4.4 & 3.4  & 2.4  & 83.6 & -4.3 & 3.4  & 2.5  & 83.7 \\[0.6em]
    \multicolumn{1}{l}{$X_2$} & Proposed (known) & 0.0  & 6.1  & 5.8  & 92.4 & 0.0  & 6.2  & 5.8  & 92.3 & -0.1 & 3.6  & 3.5  & 93.9 & -0.1 & 3.6  & 3.5  & 93.9 \\
         & Proposed (estimated) & 0.0  & 6.1  & 5.8  & 92.4 & 0.0  & 6.1  & 5.8  & 92.3 & -0.1 & 3.6  & 3.5  & 93.9 & -0.1 & 3.6  & 3.5  & 93.9 \\
         & Direct-DML (2) & 0.5  & 6.1  & 5.5  & 91.4 & 1.7  & 6.7  & 6.0  & 91.2 & 0.1  & 3.6  & 3.5  & 93.8 & 0.7  & 3.9  & 3.8  & 93.6 \\
         & Direct-DML (5) & 0.9  & 6.2  & 5.4  & 90.1 & 2.8  & 6.8  & 5.8  & 89.8 & 0.2  & 3.6  & 3.4  & 93.4 & 0.9  & 3.9  & 3.7  & 93.1 \\
         & No-DML & 0.2  & 5.5  & 4.0  & 84.8 & 0.2  & 5.5  & 4.0  & 84.9 & 0.0  & 3.2  & 2.4  & 85.9 & 0.0  & 3.2  & 2.4  & 85.9 \\
    \bottomrule
    \end{tabular}%
\end{adjustbox}
}
\footnotesize \textcolor{black}{(Bias $\times 10^{-3}$): Average bias across all simulation replicates and time points; (SD $\times 10^{-2}$): Average Monte Carlo standard deviation across all time points; (SE $\times 10^{-2}$): Average estimated standard error across all simulation replicates and time points; (CP $\%$): Average coverage probability of the 95\% confidence intervals across all time points.}
\end{table}

To implement the proposed method, we considered both the case when the carryover discounting factor was known (denoted as ``Proposed (known)"), and the case when  it was estimated as described in Remark 1 (denoted as ``Proposed (estimated)").
We compared our method with two competing approaches. The first method assumed the same carryover structure as in Equation \eqref{equ:yt0_revised} but did not use DML as in Equation \eqref{eq:DML_revised} for estimating $\beta_{0t}$ (i.e., it does not apply centering); this method is denoted as ``No-DML". The second method incorporated DML but disregarded the carryover treatment effect structure, that is, it directly modeled $h_{0t}(H_{it})$ in Equation \eqref{eq:ht_def} using some machine learning models (denoted as ``Direct-DML"). 
Incorporating the full history is likely to lead to overfitting, especially when the sample size is small. To mitigate this risk, we assumed that only the current time point and the $t_0$ most recent time points are used in all ML models.
We considered two choices for the memory length $t_0$, specifically $t_0 \in \{2, 5\}$, and denote these as Direct-DML ($t_0$).
In Case I, all functions to be estimated are (generalized) linear. Thus, we used the \textit{glmnet} package in \textit{R} \citep{friedman2010regularization} to fit the models across all scenarios. 
{In Case II, both the propensity score and baseline outcome functions exhibit a tree structure; accordingly, we use Classification and Regression Trees \citep[CART;][]{breiman1996bagging} across all methods to model these components. For the proposed and No-DML methods, CART is also applied to model the prognostic function.
For the Direct-DML methods, we explore various machine learning models (\textit{glmnet}, CART, k-Nearest Neighbors, and Random Forest) using the \textit{caret} package in \textit{R} and fine-tune hyperparameters for optimal performance. }
 Model performance was assessed using four metrics: (i) the average bias of $\wh\beta_t$ across all simulation replicates, (ii) the Monte Carlo standard deviation (of the $\wh\beta_t$), (iii) the average estimated standard error, calculated using \eqref{equ:CI}, and (iv) the coverage probabilities of the $95\%$ confidence intervals.

The simulation results based on 500 replicates are presented in Table \ref{table_simu1}. As shown, the proposed method provides estimates with minimal bias and coverage probabilities close to the nominal $95\%$ level. As the sample size increases, the coverage rate of the proposed method improves from approximately $92\sim 93\%$ to around $94 \sim 95\%$, supporting the validity of our theoretical framework. Note that even without knowledge of the true carryover discounting factor, the model accurately estimates it, yielding results nearly identical to those obtained when the carryover discounting factor is known.
In contrast, the No-DML method exhibits notable bias and fails to provide reliable inference, particularly for the intercept effects.
The Direct-DML methods perform better in bias and coverage probabilities over the No-DML method; however, it remains biased and generally falls below the nominal coverage level, especially when $n=200$ at $t_0 = 5$. This may be due to overfitting of the prognostic function $\mu_{0}(U_{it})$.
Figure S.1 in the Supplementary Materials provides a detailed view of the model performance for the proposed method with a known carryover discounting factor in Case I, displaying the average bias and coverage probabilities across all time points. The results indicate that the estimates are unbiased and the inference is accurate across all time points, with increased sample sizes leading to more precise inference.

\begin{table}[h!]
\caption{Summary of the estimated $\wh\beta_t$ across 500 simulations for Case II.} 
\label{table_simu2}
{\centering
\begin{adjustbox}{max width=\textwidth}
    \begin{tabular}{rlrcccrcccrcccrccc}
    \toprule 
     &      & \multicolumn{8}{c}{$n=200$}                           & \multicolumn{8}{c}{$n=500$} \\
     &      & \multicolumn{4}{c}{$\gamma=0.3$} & \multicolumn{4}{c}{$\gamma=0.7$} & \multicolumn{4}{c}{$\gamma=0.3$} & \multicolumn{4}{c}{$\gamma=0.7$} \\
    \multicolumn{1}{l}{$\wh\beta_t$} & \textbf{Method} & \multicolumn{1}{c}{Bias} & SD   & SE   & CP   & \multicolumn{1}{c}{Bias} & SD   & SE   & CP   & \multicolumn{1}{c}{Bias} & SD   & SE   & CP   & \multicolumn{1}{c}{Bias} & SD   & SE   & CP \\
\cmidrule(lr){3-6}\cmidrule(lr){7-10}\cmidrule(lr){11-14}\cmidrule(lr){15-18}
\multicolumn{1}{l}{Intercept} & Proposed (known) & -4.9 & 6.7  & 6.6  & 93.5 & -4.8 & 6.8  & 6.7  & 93.5 & -2.6 & 3.8  & 3.7  & 93.9 & -2.6 & 3.8  & 3.7  & 94.0 \\
         & Proposed (estimated) & -5.1 & 6.6  & 6.5  & 93.4 & -5.2 & 6.7  & 6.6  & 93.5 & -2.6 & 3.8  & 3.7  & 93.9 & -2.7 & 3.8  & 3.7  & 94.0 \\
         & Direct-DML (2) & -5.4 & 8.2  & 7.6  & 92.1 & -1.8 & 8.6  & 8.0  & 92.3 & -1.9 & 4.9  & 4.8  & 94.1 & -1.1 & 5.1  & 5.0  & 94.2 \\
         & Direct-DML (5) & -10.8 & 8.5  & 7.5  & 90.5 & -4.1 & 8.9  & 7.8  & 90.6 & -3.7 & 4.9  & 4.7  & 93.6 & -2.0 & 5.1  & 4.9  & 93.7 \\
         & No-DML & -68.7 & 8.7  & 4.1  & 49.4 & -70.3 & 8.6  & 4.1  & 48.5 & -24.4 & 4.9  & 2.4  & 66.0 & -25.4 & 4.9  & 2.4  & 66.0 \\[0.6em]
    \multicolumn{1}{l}{$Z_1$} & Proposed (known) & -0.4 & 6.8  & 6.5  & 94.8 & -0.5 & 7.0  & 6.7  & 94.7 & -0.1 & 3.7  & 3.6  & 94.9 & 0.0  & 3.7  & 3.7  & 94.9 \\
         & Proposed (estimated) & -0.4 & 6.7  & 6.4  & 94.8 & -0.4 & 6.9  & 6.6  & 94.6 & -0.1 & 3.7  & 3.6  & 94.9 & -0.1 & 3.7  & 3.7  & 94.9 \\
         & Direct-DML (2) & -0.7 & 8.0  & 7.6  & 93.1 & -2.4 & 8.5  & 8.0  & 92.7 & -0.2 & 4.8  & 4.7  & 94.5 & -1.0 & 5.1  & 4.9  & 94.1 \\
         & Direct-DML (5) & -1.2 & 8.1  & 7.4  & 92.2 & -5.4 & 8.6  & 7.7  & 91.4 & -0.4 & 4.8  & 4.7  & 94.3 & -2.0 & 5.1  & 4.9  & 93.7 \\
         & No-DML & 0.0  & 4.5  & 4.1  & 92.1 & 0.0  & 4.3  & 4.1  & 94.0 & -0.1 & 2.7  & 2.4  & 91.2 & 0.0  & 2.5  & 2.4  & 93.8 \\[0.6em]
    \multicolumn{1}{l}{$Z_2$} & Proposed (known) & -0.4 & 6.9  & 6.6  & 94.8 & -0.4 & 7.1  & 6.8  & 94.7 & 0.1  & 3.8  & 3.7  & 94.8 & 0.1  & 3.8  & 3.7  & 94.9 \\
         & Proposed (estimated) & -0.5 & 6.8  & 6.5  & 94.7 & -0.6 & 7.0  & 6.7  & 94.7 & 0.0  & 3.7  & 3.7  & 94.8 & 0.1  & 3.8  & 3.7  & 94.8 \\
         & Direct-DML (2) & 0.5  & 8.0  & 7.6  & 92.6 & 2.4  & 8.5  & 8.0  & 92.4 & 0.3  & 4.9  & 4.8  & 94.1 & 1.1  & 5.1  & 5.0  & 93.7 \\
         & Direct-DML (5) & 0.9  & 8.2  & 7.4  & 91.4 & 5.1  & 8.6  & 7.8  & 90.9 & 0.5  & 4.9  & 4.7  & 93.9 & 2.1  & 5.2  & 4.9  & 93.4 \\
         & No-DML & -1.2 & 4.4  & 4.1  & 93.2 & -0.7 & 4.3  & 4.1  & 94.0 & 0.6  & 2.4  & 2.4  & 94.5 & 0.3  & 2.4  & 2.4  & 94.6 \\[0.6em]
    \multicolumn{1}{l}{$X_1$} & Proposed (known) & 0.0  & 7.5  & 7.3  & 94.6 & 0.2  & 7.6  & 7.4  & 94.5 & -2.3 & 4.1  & 4.0  & 94.4 & -2.2 & 4.1  & 4.1  & 94.5 \\
         & Proposed (estimated) & -0.4 & 7.3  & 7.2  & 94.4 & -0.2 & 7.5  & 7.3  & 94.5 & -2.4 & 4.0  & 4.0  & 94.5 & -2.2 & 4.1  & 4.1  & 94.5 \\
         & Direct-DML (2) & -2.1 & 8.7  & 8.2  & 92.2 & -2.4 & 9.1  & 8.5  & 92.0 & -0.9 & 5.3  & 5.1  & 93.9 & -1.2 & 5.5  & 5.4  & 93.8 \\
         & Direct-DML (5) & -4.4 & 8.8  & 7.9  & 91.0 & -4.8 & 9.3  & 8.3  & 90.8 & -1.7 & 5.3  & 5.1  & 93.5 & -2.2 & 5.5  & 5.3  & 93.4 \\
         & No-DML & -31.8 & 5.1  & 4.2  & 82.2 & -24.4 & 5.0  & 4.2  & 86.1 & -29.2 & 3.9  & 2.4  & 59.0 & -24.3 & 3.8  & 2.4  & 66.4 \\[0.6em]
    \multicolumn{1}{l}{$X_2$} & Proposed (known) & -0.3 & 7.4  & 7.1  & 94.6 & -0.2 & 7.6  & 7.3  & 94.4 & 0.0  & 4.1  & 3.9  & 94.6 & 0.0  & 4.1  & 4.0  & 94.6 \\
         & Proposed (estimated) & -0.2 & 7.3  & 7.0  & 94.5 & -0.2 & 7.5  & 7.2  & 94.5 & 0.0  & 4.0  & 3.9  & 94.6 & -0.1 & 4.1  & 4.0  & 94.6 \\
         & Direct-DML (2) & -0.1 & 8.6  & 8.1  & 92.3 & 0.1  & 9.1  & 8.5  & 92.2 & -0.1 & 5.2  & 5.1  & 93.9 & 0.0  & 5.5  & 5.3  & 93.9 \\
         & Direct-DML (5) & -0.2 & 8.8  & 8.0  & 91.3 & 0.1  & 9.2  & 8.3  & 91.1 & -0.1 & 5.3  & 5.0  & 93.5 & 0.1  & 5.5  & 5.3  & 93.5 \\
         & No-DML & 0.0  & 4.7  & 4.4  & 93.4 & -0.1 & 4.7  & 4.4  & 93.4 & 0.0  & 2.7  & 2.5  & 93.8 & 0.0  & 2.7  & 2.5  & 93.7 \\
    \bottomrule
    \end{tabular}%
\end{adjustbox}
}
\footnotesize \textcolor{black}{(Bias $\times 10^{-3}$): Average bias across all simulation replicates and time points; (SD $\times 10^{-2}$): Average Monte Carlo standard deviation across all time points; (SE $\times 10^{-2}$): Average estimated standard error across all simulation replicates and time points; (CP $\%$): Average coverage probability of the 95\% confidence intervals across all time points.}
\end{table}

For Case II, 
as shown in Table \ref{table_simu2}, the No-DML method exhibits substantial bias and fails to provide reliable inference, achieving a coverage rate of only around $50\%$ for the intercept effects when $n = 200$. This issue likely arises from the high correlation between the propensity score and the prognostic function, as discussed in \cite{chernozhukov2018double}.
In the Direct-DML methods, the limited ability of 
a single ML model to capture both the nonlinear tree structure of the prognostic function and the linear carryover treatment effects results in higher standard errors for the estimated $\wh\beta_t$. Consequently, this leads to wider confidence intervals and reduced statistical power compared to the proposed method.
Overall, the proposed method demonstrates superior performance across all evaluated metrics.

\section{Application}
\label{sec:application}

We applied our method to the Mobile Parkinson’s Observatory for Worldwide Evidence-based Research (mPower) study \citep{bot2016mpower}. The mPower  was a six-month longitudinal digital health observational study designed to assess the feasibility of smartphone-based remote monitoring of Parkinson’s disease (PD) symptoms. Its broader objective was to develop digital biomarkers for use in future clinical drug trials.
The study included 1,087 participants who self-identified as having a professional PD diagnosis. Among these participants, the majority were taking daily levodopa treatment. While not mandatory, participants were encouraged to complete up to three daily activities throughout the study period. These activities involved five active assessments (tapping, voice, walking, balance, and memory), during which smartphone sensors recorded relevant data. 

{\color{black} In this application, all subjects took levodopa every day, so for any time \(t\), the prior treatment history satisfies \(A_{is}=1\) for \(s=1,\dots,t-1\). On days when the phenotype was recorded immediately after levodopa administration, the observed outcome corresponds to \(Y_{it}(\bar A_{i,t-1},1)\). On days when the phenotype was measured before the levodopa dose on day \(t\), the observed outcome corresponds to \(Y_{it}(\bar A_{i,t-1},0)\), assuming that the potential outcome just before treatment is the same as the potential outcome that would have been observed had the treatment at time \(t\) not been given. The schedule of measuring phenotypes before or after medication varied across subjects.}

In our analysis, we mainly focus on the tapping modality. The tapping assessment evaluates bradykinesia (slowness of movement), which is one of the hallmark symptoms of PD. During the assessment, participants placed the phone on a flat surface and used two fingers from the same hand to alternately tap two fixed points on the screen for 20 seconds. The software recorded both the location and timing of each tap, providing detailed data on tapping speed and rhythm. For our analysis, we used the ``mean tapping interval" (i.e., the average time difference between two taps; smaller intervals indicate quicker and better performance) as the outcome variable and applied a logarithmic transformation to address its skewness.
We included baseline demographic variables, age, sex, smoking history, years since diagnosis of PD, and deep brain stimulation status (whether a patient has undergone the procedure), to investigate potential heterogeneity in treatment effects based on these factors.
We did not include any time-varying exogenous variables. Instead, we incorporated features derived from the walking and resting modalities as prognostic variables. These modalities capture aspects of tremor, motor fluctuations, and gait disturbances during walking or standing still, and were included to improve adjustment for participants' time-varying clinical status.
A more detailed definition of these features is available in \cite{xu2023mixed} and \cite{omberg2022remote}.

\begin{figure}[h]
    \centering
    \includegraphics[width=0.9\textwidth]{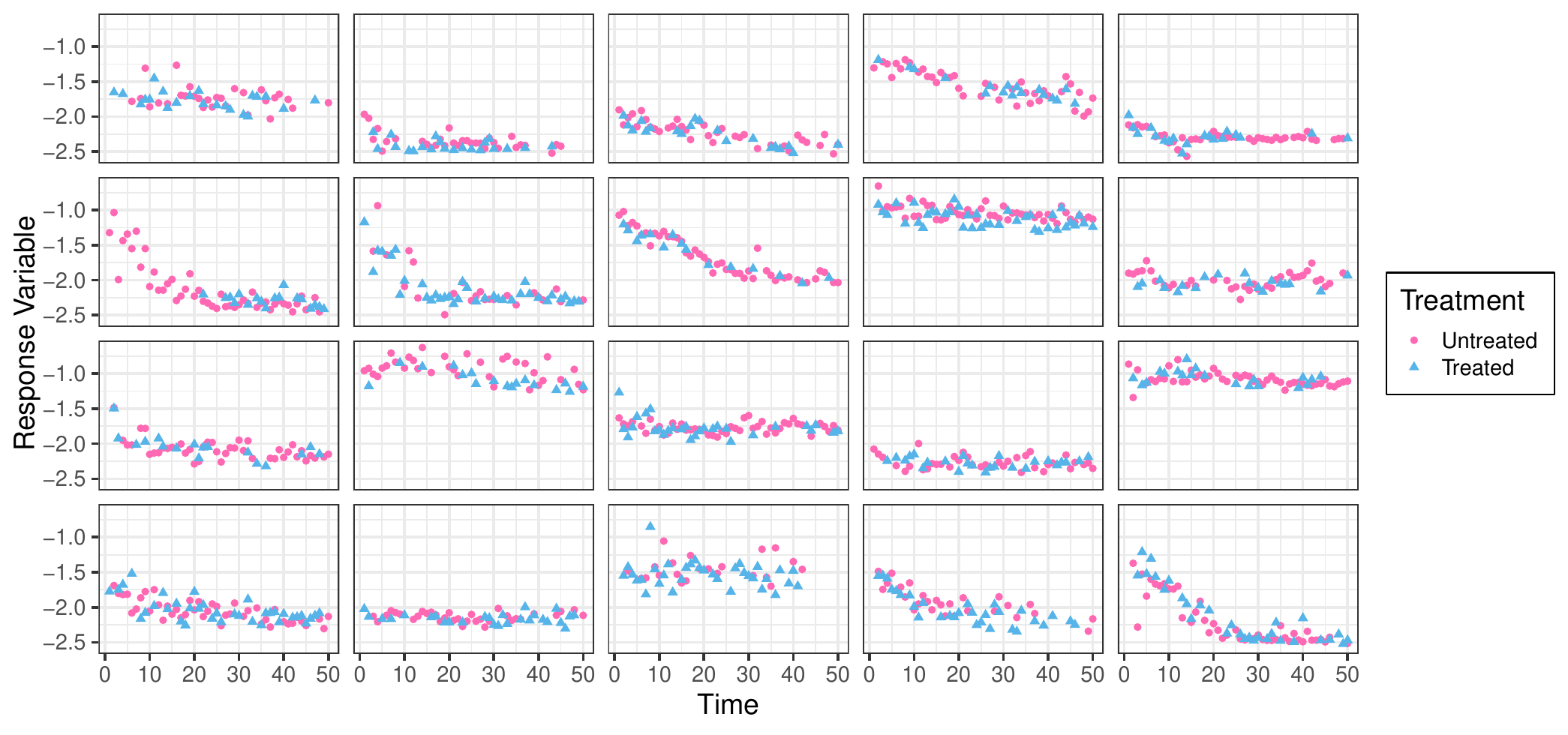}
    \caption{Scatterplot of the response variable (mean tapping intervals) for 20 randomly selected subjects from the mPower data.}
    \label{img_real_Y}
\end{figure}

Due to the presence of noisy sensor data, preprocessing was necessary. We followed the preprocessing methods outlined in \cite{xu2023mixed}.
Our analysis revealed that most patients had at most one tapping measurement per day. 
Since the afternoon data were significantly more complete and exhibited different patterns compared to the morning data due to akinesia \citep{xu2023mixed,omberg2022remote}, we restricted our analysis to only measurements of the afternoon.
To ensure data quality, we excluded subjects with excessive missing tapping or walking/resting measurements. After this exclusion, $283$ subjects remained in our final analysis. 
{To handle missing time-varying covariates, we used the ``\textit{mice}" procedure in R \citep{van2011mice} for imputation. Specifically, we treated each patient's digital phenotypes as a cluster, used the day 
as the random effects, and included both the baseline and time-varying covariates in the imputation.}
Figure \ref{img_real_Y} presents the scatterplots of the mean tapping intervals for $20$ randomly selected subjects over $50$ days by treatment status (treatment status was distinguished using different colors).

\begin{figure}[h]
    \centering
    \includegraphics[width=0.63\textwidth]{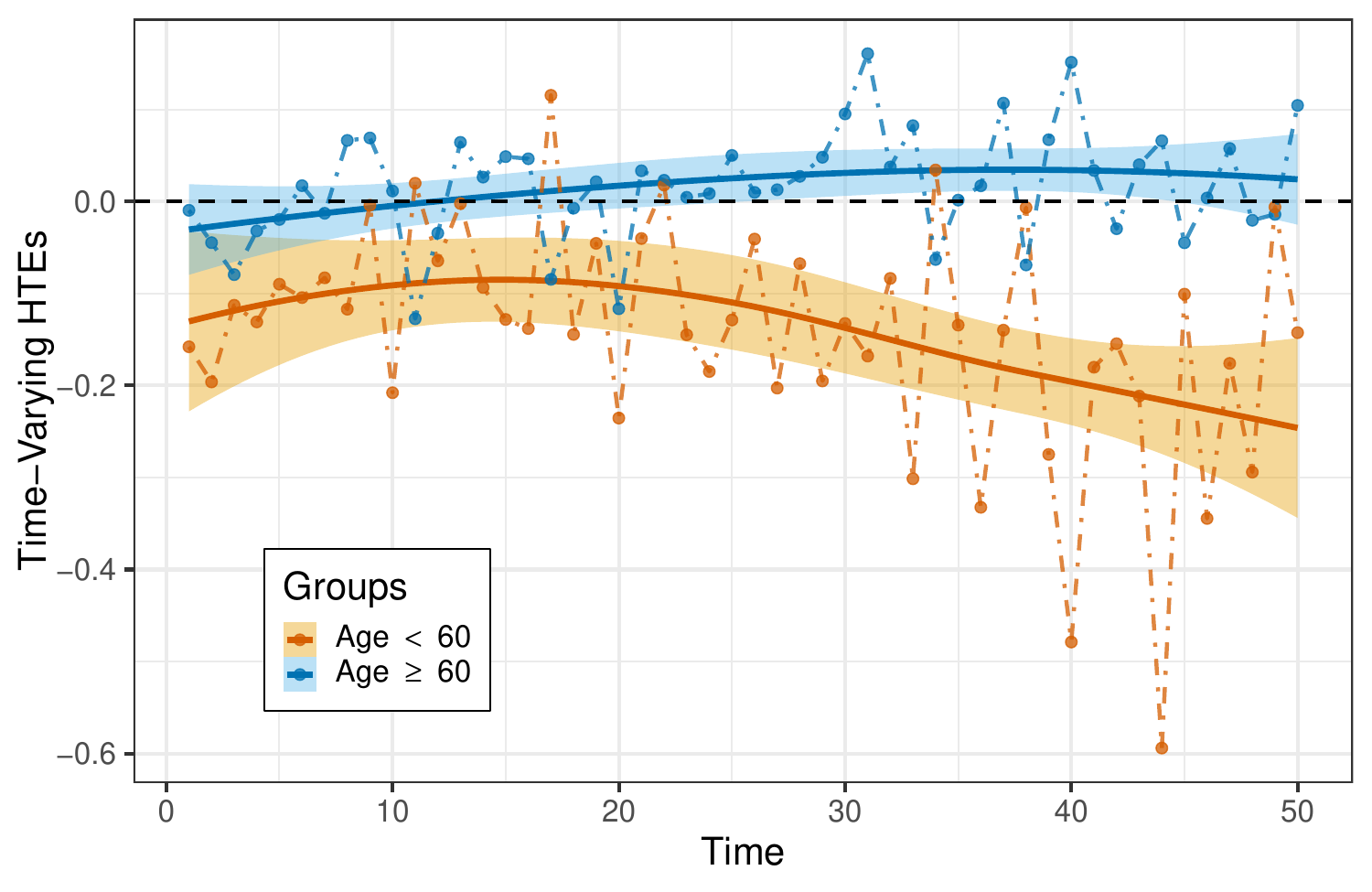}
    \caption{Visualization of the time-varying age-specific HTE in the mPower data. Solid dots represent the estimated treatment effect, dashed lines illustrate the time-varying trends, smoothed solid curves show the fitted smoothed trends of the time-varying HTE, and shaded areas indicate the $95\%$ confidence bands for the mean estimation.}
    \label{fig_real}
\end{figure}

We applied our proposed method to analyze the time-varying HTEs over the first $50$ days. 
For this analysis, a negative value of treatment effect indicates an improvement in treatment outcomes, as smaller values of the mean tapping interval represent better fine motor function.
Our findings suggested that the treatment was more effective for younger PD patients compared to older patients. Specifically, we dichotomized age at its median for better interpretability (i.e., younger than $60$ years of age versus 60 or older). The time-varying HTEs for these two age groups are presented in Figure \ref{fig_real}(a). From the figure, it is evident that there is no consistent treatment effect for patients aged $60$ or older, while a clear treatment effect is observed for younger patients.
To smooth the time-varying effects, we applied a local smoothing method \citep{cleveland1979robust} and visualized the mean curve with a $95\%$ confidence band (represented by the shaded area) using ``\textit{loess}" in R. The results show an increasing trend in the treatment effect size from day one to day $50$.


\begin{figure}[h]
    \centering
    \begin{subfigure}[b]{0.50\textwidth}
        \centering
        \includegraphics[width=\textwidth]{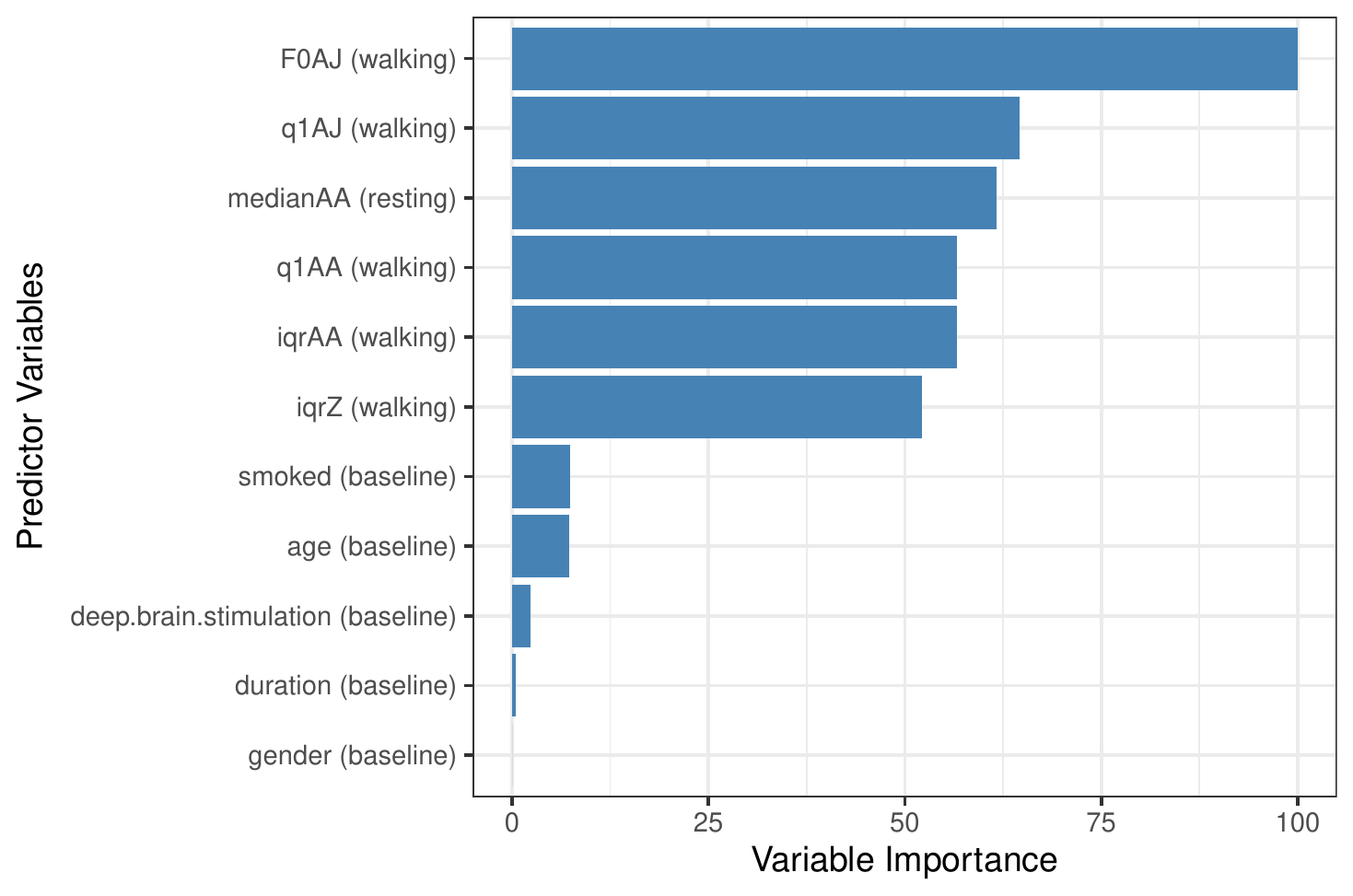}
        \caption{Variable importance for the propensity scores}
    \end{subfigure}
    \hfill
    \begin{subfigure}[b]{0.47\textwidth}
        \centering
        \includegraphics[width=\textwidth]{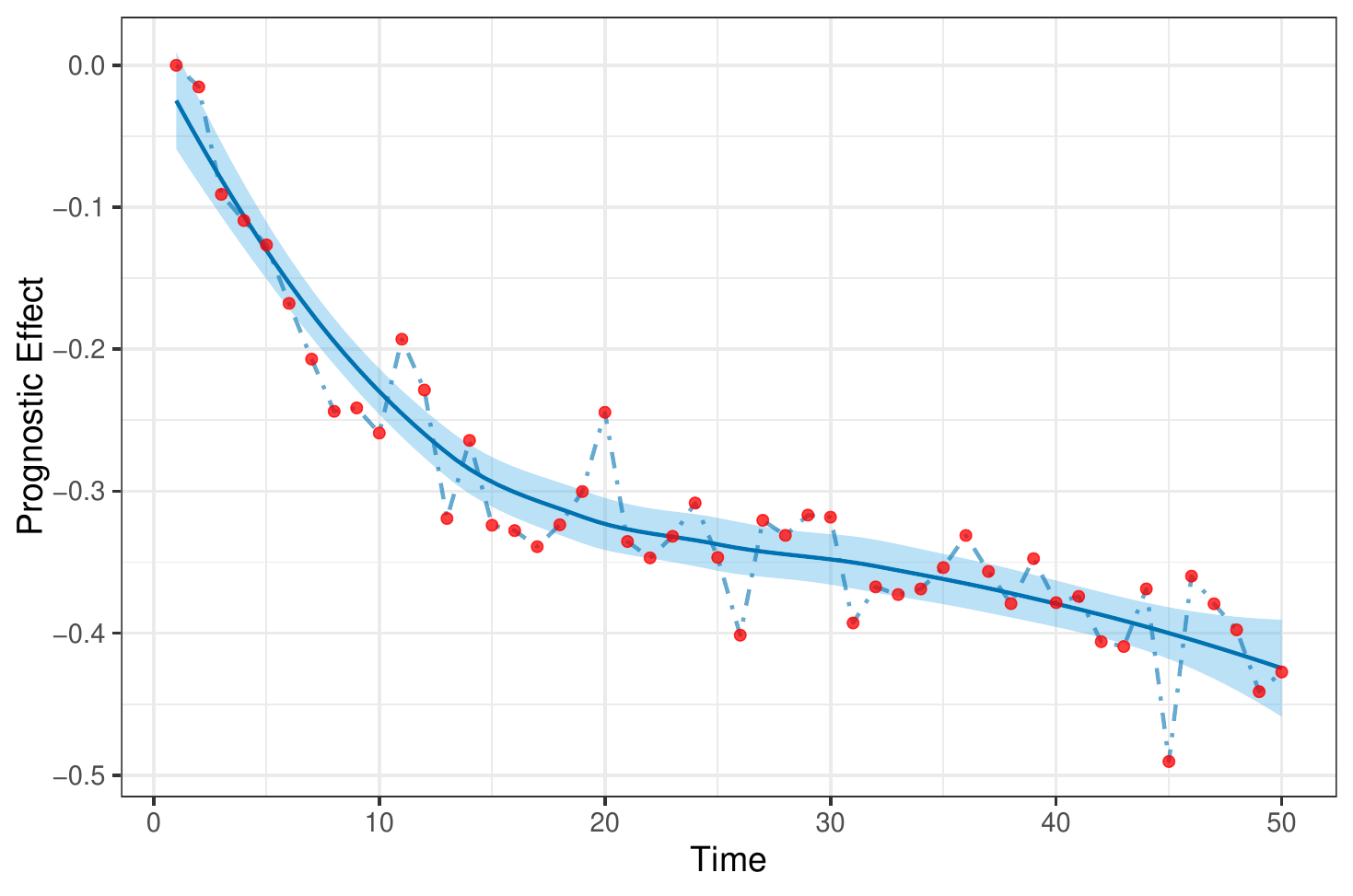}
        \caption{Prognostic effect over time}
    \end{subfigure}
    \caption{Details of the propensity score and prognostic effect models.}
    \label{fig_real2}
\end{figure}

For model fitting, we estimated the carryover discounting factor as $0.1$ using a grid search over the range $\{0, 0.1, \ldots, 0.9, 1\}$, minimizing the RMSE. This result suggests that the carryover treatment effect in this application is relatively small, which is expected given the short half-life of levodopa medication \citep{contin2010pharmacokinetics}.
To estimate the propensity score $\pi_{0t}(U_{it})$, we trained a Gradient Boosting Machine (GBM) model using the \textit{caret} package in R. The relative importance of variables \citep{friedman2001greedy} is presented in Figure \ref{fig_real2}(a). Detailed descriptions of each variable can be found at {\url{https://github.com/Sage-Bionetworks/mpowertools/blob/master/FeatureDefinitions.md}}. The results suggest that incorporating walking/resting modalities increases the accuracy of the propensity score model. 
Figure \ref{fig_real2}(b) illustrates the main prognostic effects over time, revealing a clear decreasing trend. This pattern aligns with the patterns of the outcome variable (tapping intervals) in Figure \ref{img_real_Y}. A possible explanation is that as time progresses, participants become more familiar with the task, leading to improved overall performance similar to a practice effect. This analysis shows that it is important to separate the prognostic effect from the treatment effect. 

\vspace{-1em}
\section{Discussion}
\label{sec:discussion}
\vspace{-0.75em}

In this paper, we proposed a DML method to estimate the time-varying heterogeneous treatment effects (HTEs) using digital phenotype data. By decomposing the treatment effect on the outcome into three distinct components (baseline effect, the cumulative direct effects from interventions, and additional prognostic effect), our approach achieved a more accurate estimation of heterogeneous HTEs and captures how the treatment effect evolves over time.
We used a sequential procedure combined with DML to model both time-varying instantaneous and carryover treatment effects. 
We proved the asymptotic normality of the proposed estimator, facilitating robust inference.
Extensive simulation studies validated the finite-sample performance of our method and demonstrated the advantages of utilizing DML and decomposing treatment effects compared to a regular DML.
Applying our approach to the mPower study, we found that the treatment was more effective for younger PD patients than for older ones. 
We also found a significant prognostic effect consistent with the practice effect.  These findings highlight the utility of the proposed method.

In mHealth studies, causal excursion effect \citep{boruvka2018assessing, qian2020linear} measures the impact of a ``one-step deviation" from a baseline policy or treatment sequence, capturing the short-term causal effect of an intervention at a given time. These effects are useful for evaluating just-in-time adaptive interventions (JITAIs) in mHealth studies, where treatments are delivered dynamically based on contextual information (e.g., digital phenotypes such as physical activity, mood, or physiological states). Our method can be extended to estimate the causal excursion effect when the treatment administration differs from the observed treatment sequences based on the proposed models. In addition, beyond its application in PD, our proposed method could be applied to track and optimize insulin management in diabetes to maintain glycemic control or monitor walking patterns and gait stability in older populations, identify early signs of decline, and tailor interventions to maintain mobility and prevent falls.

In Section \ref{sec:theory}, the Donsker conditions in Condition 4 may be further relaxed through data splitting \citep{chernozhukov2018double}. Doing so removes the need for the Donsker assumption and permits the use of more flexible machine learning estimators, as long as they satisfy the rate conditions in Condition 3. In addition, data splitting is known to mitigate overfitting. However, because the target parameter depends on the evolving treatment and covariate history, extending standard data-splitting arguments to this dynamic setting is nontrivial. For this reason, we do not use data splitting in the paper.

\section*{Acknowledgements}

This research is supported by U.S. National Institutes of Health grants  MH123487, NS073671 and GM124104.






\vspace{-2em}



\bibliographystyle{apalike}
\bibliography{ref}


\clearpage\pagebreak\newpage
\setcounter{section}{0}
\renewcommand{\thesection}{S.\arabic{section}} 
\renewcommand{\thesubsection}{S.\arabic{section}.\arabic{subsection}}
\setcounter{page}{1}
\renewcommand{\theequation}{S.\arabic{equation}}
\renewcommand{\thefigure}{S.\arabic{figure}}
\renewcommand{\thetable}{S.\arabic{table}}
\setcounter{table}{0}
\setcounter{equation}{0} \setcounter{figure}{0} 
		
\renewcommand{\theenumi}{\arabic{enumi}}

\begin{center}
{\Large{\bf Supplementary Materials for ``Cumulative Marginal Mean Model for Assessing
Sequential Effects Using Digital Health Data"}}\\ 
			\vskip1cm
   Xingche Guo, Zexi Cai, Yuanjia Wang, and Donglin Zeng
\end{center}

\vskip1cm

\section*{Appendix A}

The Appendix A contains the proof of theorems. First, note that for the little-$o$ notation, $ {x}_n = o(r_n)$ for $ {x}_n \in \mathbb{R}^d$ means that $ {a}^{\top}  {x}_n = o_p(r_n)$ for all fixed $ {a} \in \mathbb{R}^d$, and $ {A}_n = o(r_n)$ for $ {A}_n \in \mathbb{R}^{p \times q}$ means that $ {b}^{\top} \cdot \mathrm{vec}( {A}_n) = o(r_n)$ for all fixed $ {b} \in \mathbb{R}^{pq}$. 
It is straightforward to show that if $ {A}_n$ is positive definite, then $ {A}_n = o_p(r_n)$ is equivalent to $\Lambda_{\max}( {A}_n) = o_p(r_n)$.
Similar arguments hold for big-$O$ notation.

\bigskip

\noindent\textbf{Proof of Proposition 1.}

Define 
\begin{align*}
    C_t = C_t(Z, \bar A_{t-1}, \bar X_{t-1}) = \sum_{k=1}^{t-1} A_{k}\, \gamma^{t-k}\,\beta_{0k}^\top \widetilde{X}_{k}
\end{align*}
We have
\begin{eqnarray*}
& &E(R_{t}\mid U_{t},A_{t}=1)-E(R_{t}\mid U_{t},A_{t}=0)  \\
&=&E(Y_{t}-
C_t \mid U_{t},A_{t}=1) - \beta_{0t}^{\top} \widetilde X_{it} -E(Y_{it}-
C_t \mid U_{t},A_{t}=0) \\
&=&E \left[ E(Y_{t}(\bar A_{t-1},1)-
C_t|{ H}_{t}, A_{t}=1)\mid U_{t}, A_{t}=1 \right] - \beta_{0t}^{\top} \widetilde X_{it}\\
& &-
E \left[E(Y_{t}(\bar A_{t-1},0)-
C_t|{ H}_{t}, A_{it}=0)\mid U_{t}, A_{t}=0 \right]  \ \ \textrm{(by Condition 1a)}\\
&=& E\left[ E(Y_{t}(\bar A_{t-1},1)-
C_t|{H}_{t})\mid U_{t}, A_{t}=1 \right] - \beta_{0t}^{\top} \widetilde X_{it} \\
& &-
E \left[ E(Y_{t}(\bar A_{t-1},0)-
C_t|{H}_{t})\mid U_{t}, A_{t}=0 \right]  \ \ \textrm{(by Condition 1b)}\\
&=&E \left[ E(Y_{t}(\bar A_{t-1},1)-
C_t|{ H}_{it})\mid U_{t} \right] -\beta_t\widetilde X_{t} \\
& &-
E \left[ E(Y_{t}(\bar A_{t-1},0)-
C_t|{ H}_{t})\mid U_{t} \right] \ \ \textrm{(by Condition 1d)}\\
&=&E \left[Y_{t}(\bar A_{t-1},1) -
Y_{t}(\bar A_{t-1},0)\mid U_{t} \right]-\beta_{0t}^{\top}\widetilde X_{t} \\
&=&E \Big[ E\left[Y_{t}(\bar A_{t-1},1)- Y_{t}(\bar A_{t-1},0)\mid \bar A_{t-1}, \bar X_{t-1}, U_{t} \right]\mid U_{it} \Big] - \beta_{0t}^{\top}\widetilde X_{t}\\
&=&E \Big[ E\left[Y_{t}(\bar A_{t-1},1)- Y_{t}(\bar A_{t-1},0)\mid Z, \bar A_{t-1}, \bar X_{t-1}, X_t \right]\mid U_{it} \Big] - \beta_{0t}^{\top} \widetilde X_{t} \ \ \textrm{(by Condition 1c)}\\
&=&E(\beta_{0t}^{\top} \widetilde X_{t}\mid U_{t}) -\beta_{0t}^{\top}\widetilde X_{t} \ \ \textrm{by Equation (2.2)}\\
&=&0.
\end{eqnarray*}

\noindent\textbf{Proof of Proposition 2.}
Evaluating the score at the true parameter \((\beta_{0t},\eta_{0t})\) gives
\[
S_t(\beta_{0t},\eta_{0t})
=
\Bigl\{
Y_t-A_t\beta_{0t}^\top \widetilde X_t-h_{0t}(H_t)
\Bigr\}
\Bigl\{
A_t-\pi_{0t}(U_t)
\Bigr\}\widetilde X_t
=
W_tV_t\widetilde X_t.
\]
Define deviation 
\begin{align*}
    W_{t} = Y_t - A_t \beta_{0t}^\top \widetilde{X}_t - h_{0t}(H_t) \qquad \mbox{and}\qquad V_t = A_t - \pi_{0t}(U_t)
\end{align*}
Hence,
\[
E\{S_t(\beta_{0t},\eta_{0t})\}
=
E\!\left[W_tV_t\widetilde X_t\right].
\]
By iterated expectation conditioning on \(U_t\), and using that \(\widetilde X_t\) is measurable with respect to \(U_t\),
\[
E\!\left[W_tV_t\widetilde X_t\right]
=
E\!\left[
E(W_tV_t\mid U_t)\widetilde X_t
\right].
\]
Now,
\[
E(W_tV_t\mid U_t)
=
E\!\left[
W_t\{A_t-\pi_{0t}(U_t)\}\mid U_t
\right].
\]
Applying iterated expectation again,
\[
E(W_tV_t\mid U_t)
=
E\!\left[
E(W_t\mid A_t,U_t)\{A_t-\pi_{0t}(U_t)\}\mid U_t
\right].
\]
By Proposition 1 and the fact that $W_t = R_t - \mu_{0t}(U_t)$., 
\begin{align*}
     E(W_t\mid A_t,U_t) 
    = E(R_t\mid A_t, U_t) - \mu_{0t}(U_t) 
    = E(R_t\mid U_t) - \mu_{0t}(U_t) 
    = 0.
\end{align*}
Hence $E(W_tV_t\mid U_t)=0$ and $E\{S_t(\beta_{0t},\eta_{0t})\}=0$,
which proves part (i).

Next, consider a local perturbation of the nuisance parameter of the form $\eta_{0t}+\rho\,\Delta\eta_t$.
At \(\beta_t=\beta_{0t}\), we have
\[
S_t(\beta_{0t},\eta_{0t}+\rho\,\Delta\eta_t)
=
\bigl\{W_t-\rho\,\Delta h_t(H_t)\bigr\}
\bigl\{V_t-\rho\,\Delta\pi_t(U_t)\bigr\}
\widetilde X_t.
\]
Differentiating with respect to \(\rho\) and evaluating at \(\rho=0\) yields
\begin{align*}
\left.
\frac{\partial}{\partial \rho}
E\bigl[S_t(\beta_{0t},\eta_{0t}+\rho\,\Delta\eta_t)\bigr]
\right|_{\rho=0}
&=
-
E\!\left[
\bigl\{
V_t\,\Delta h_t(H_t)
+
W_t\,\Delta\pi_t(U_t)
\bigr\}\widetilde X_t
\right].
\end{align*}
Thus, it suffices to show that both expectations vanish.

For the second term, by iterated expectation,
\begin{align*}
E\!\left[W_t\,\Delta\pi_t(U_t)\,\widetilde X_t\right]
&=
E\!\left[
E(W_t\mid U_t)\,\Delta\pi_t(U_t)\,\widetilde X_t
\right] = 0.
\end{align*}

For the first term, again by iterated expectation,
\begin{align*}
E\!\left[V_t\,\Delta h_t(H_t)\,\widetilde X_t\right]
&=
E\!\left[
E(V_t\mid H_t)\,\Delta h_t(H_t)\,\widetilde X_t
\right].
\end{align*}
By Condition 1(d) and the definition of \(V_t\),
\[
E(V_t\mid H_t)
=
E\!\left[A_t-\pi_{0t}(U_t)\mid H_t\right]
=
P(A_t=1\mid H_t)-P(A_t=1\mid U_t) = 0.
\]
Therefore $E\!\left[V_t\,\Delta h_t(H_t)\,\widetilde X_t\right] = 0$ and hence
\[
\left.
\frac{\partial}{\partial \rho}
\Psi_t\bigl(\beta_{0t},\eta_{0t}+\rho\,\Delta\eta_t\bigr)
\right|_{\rho=0}
=0,
\]
which establishes part (ii).
\hfill $\square$

\vspace{2em}
\noindent\textbf{Proof of Theorem 1}:
For each $t\in\mathcal T$, define
\[
J_{0t}
=
E\!\left[
A_t\{A_t-\pi_{0t}(U_t)\}\widetilde X_t\widetilde X_t^\top
\right],
\]
and recall that 
\[
\widehat J_t=\Psi_t^a(\widehat\eta_t),
\qquad
\widehat\eta_t=(\widehat h_t,\widehat\pi_t), \qquad  \widehat\pi_t(\cdot) = \widehat\pi(\cdot, t).
\]
Let $\bbG_n=n^{1/2}(\bbP_n-\bbP)$
denote the empirical process. By the DML estimating equation,
\[
0
=
\bbP_n S_t(\widehat\bbeta_t,\widehat\eta_t)
=
\Psi_t^b(\widehat\eta_t)-\widehat J_t\,\widehat\bbeta_t,
\]
and hence
\vspace{-1em}
\[
\widehat\bbeta_t=\widehat J_t^{-1}\Psi_t^b(\widehat\eta_t).
\]
Therefore,
\vspace{-1em}
\begin{align*}
n^{1/2}(\widehat\bbeta_t-\bbeta_{0t})
&=
n^{1/2}\widehat J_t^{-1}\Bigl\{\Psi_t^b(\widehat\eta_t)-\widehat J_t\bbeta_{0t}\Bigr\} \\
&=
\widehat J_t^{-1}\,n^{1/2}\bbP_n S_t(\bbeta_{0t},\widehat\eta_t).
\end{align*}
Next, decompose
\begin{align*}
n^{1/2}\bbP_n S_t(\bbeta_{0t},\widehat\eta_t)
&=
\bbG_n S_t(\bbeta_{0t},\widehat\eta_t)
+
n^{1/2}\bbP S_t(\bbeta_{0t},\widehat\eta_t) \\
&=
\bbG_n S_t(\bbeta_{0t},\eta_{0t})
+
\bbG_n\Bigl\{
S_t(\bbeta_{0t},\widehat\eta_t)-S_t(\bbeta_{0t},\eta_{0t})
\Bigr\}
+
n^{1/2}\bbP S_t(\bbeta_{0t},\widehat\eta_t).
\end{align*}
Define
\[
R_{2t}
=
n^{1/2}\bbP S_t(\bbeta_{0t},\widehat\eta_t),
\qquad
R_{3t}
=
\bbG_n\Bigl\{
S_t(\bbeta_{0t},\widehat\eta_t)-S_t(\bbeta_{0t},\eta_{0t})
\Bigr\}.
\]
Hence,
\begin{align}
n^{1/2}(\widehat\bbeta_t-\bbeta_{0t})
=
\widehat J_t^{-1}
\Bigl\{
\bbG_n S_t(\bbeta_{0t},\eta_{0t})+R_{2t}+R_{3t}
\Bigr\}.
\label{eq:proof_basic_expansion}
\end{align}

We now establish the orders of the three remainder terms.

\noindent\textbf{Step 1}: $\widehat J_t-J_{0t}=o_p(1)$ and $\widehat J_t^{-1}-J_{0t}^{-1}=o_p(1)$.

Define $ S_t^a(\pi_t)
=
A_t\{A_t-\pi_t(U_t)\}\widetilde X_t\widetilde X_t^\top$, we have
\begin{align*}
\widehat J_t-J_{0t}
&=
\bbP_n S_t^a(\widehat\pi_t)-\bbP S_t^a(\pi_{0t}) \\
&=
n^{-1/2}\bbG_n S_t^a(\widehat\pi_t)
+
\bbP\Bigl\{S_t^a(\widehat\pi_t)-S_t^a(\pi_{0t})\Bigr\},
\end{align*}

Consider the limiting distribution of $\bbG_n  \mathrm{vec}( \bS_t^a(\wh\pi_t) )$. Note that
\begin{align*}
    \bbP \left\{    \mathrm{vec}( \bS_t^a(\wh\pi_t) )^{\otimes 2} \right\} = \bbP A_t \left\{A_{t}-\wh{\pi}_t(U_{t}) \right\}^2 (\widetilde X_{t} \widetilde X_{t}^{\top}) \otimes (\widetilde X_{t} \widetilde X_{t}^{\top}).
\end{align*}
Based on Condition 4 and the requirement that the valid estimate $\wh\pi_t$ must lie within the range $[0, 1]$, we can infer that $\Lambda_{\max}(\bbP \left\{    \mathrm{vec}( \bS_t^a(\wh\pi_t) )^{\otimes 2} \right\}) \le \kappa_{\max}^4 < \infty$.
Under Conditions 3, the class $\mathcal S_t^a
=
\bigl\{
S_t^a(\pi_t):\pi_t \in\mathcal F_\pi
\bigr\}$
is $P$-Donsker, and therefore $\bbG_n S_t^a(\widehat\pi_t)=O_p(1).$
Hence
\[
n^{-1/2}\bbG_n S_t^a(\widehat\pi_t)=o_p(1).
\]

For the second term, let ${b}\in\mathbb R^{d^2}$ be fixed. Then, by Cauchy--Schwarz,
\begin{align*}
\Big|
{b}^\top \mathrm{vec}
\Bigl(
\bbP\{S_t^a(\widehat\pi_t)-S_t^a(\pi_{0t})\}
\Bigr)
\Big|
&=
\Big|
\bbP\Big[
A_t\{\pi_{0t}(U_t)-\widehat\pi_t(U_t)\}
\,{b}^\top\{\widetilde X_t\otimes\widetilde X_t\}
\Big]
\Big| \\
&\le
\Bigl[
\bbP\{\widehat\pi_t(U_t)-\pi_{0t}(U_t)\}^2
\Bigr]^{1/2}
\Bigl[
\bbP
\bigl\{
{b}^\top(\widetilde X_t\otimes\widetilde X_t)
\bigr\}^2
\Bigr]^{1/2}.
\end{align*}
By Condition 3, the second factor is finite uniformly in $t$, while by Condition 4 the first factor is $o_p(1)$. Thus
\[
\bbP\{S_t^a(\widehat\pi_t)-S_t^a(\pi_{0t})\}=o_p(1),
\]
and hence
\[
\widehat J_t-J_{0t}=o_p(1).
\]

Next, by Conditions 2 and 3,
\[
J_{0t}
=
E\!\left[
\pi_{0t}(U_t)\{1-\pi_{0t}(U_t)\}
\widetilde X_t\widetilde X_t^\top
\right]
\]
is positive definite, with eigenvalues uniformly bounded away from zero and infinity. Therefore \(J_{0t}^{-1}=O(1)\). Since \(\widehat J_t-J_{0t}=o_p(1)\), Weyl's inequality implies that the eigenvalues of \(\widehat J_t\) are bounded away from zero with probability tending to one, and hence
$
\widehat J_t^{-1}=O_p(1).
$
Using the matrix identity
$$\widehat J_t^{-1}-J_{0t}^{-1}
=
\widehat J_t^{-1}(J_{0t}-\widehat J_t)J_{0t}^{-1},$$
we conclude that $\widehat J_t^{-1}-J_{0t}^{-1}=o_p(1)$.

\noindent\textbf{Step 2:} $R_{2t}=o_p(1)$.

Let $\Delta\widehat\pi_t=\widehat\pi_t-\pi_{0t}$, $\Delta\widehat h_t=\widehat h_t-h_{0t}$. Define deviation 
\begin{align*}
    W_{t} = Y_t - A_t \beta_{0t}^\top \widetilde{X}_t - h_{0t}(H_t) \qquad \mbox{and}\qquad V_t = A_t - \pi_{0t}(U_t)
\end{align*}
By the decomposition of the score,
\[
S_t(\bbeta_{0t},\widehat\eta_t)
=
\Bigl\{
W_t-\Delta\widehat h_t(H_t)
\Bigr\}
\Bigl\{
V_t-\Delta\widehat\pi_t(U_t)
\Bigr\}
\widetilde X_t.
\]
Under Condition 1, and by arguments analogous to those used in the proof of Proposition 1, each of the first-order terms has mean zero. Specifically,
\begin{align*}
    \bbP\!\left(W_t V_t \widetilde X_t\right) &= 0, \\
    \bbP\!\left\{W_t \Delta\widehat\pi_t(U_t)\widetilde X_t\right\}
    &= E\!\left[
    E\!\left(W_t \mid U_t\right)\Delta\widehat\pi_t(U_t)\widetilde X_t
    \right]
    = 0, \\
    \bbP\!\left\{V_t \Delta\widehat h_t(H_t)\widetilde X_t\right\}
    &= E\!\left[
    E\!\left(V_t \mid H_t\right)\Delta\widehat h_t(H_t)\widetilde X_t
    \right]
    = 0,
\end{align*}
It follows that
\[
\bbP S_t(\bbeta_{0t},\widehat\eta_t)
=
\bbP\Bigl[
\Delta\widehat h_t(H_t)\,
\Delta\widehat\pi_t(U_t)\,
\widetilde X_t
\Bigr].
\]
Therefore, for any fixed ${a}\in\mathbb R^d$,
\begin{align*}
\Big|
{a}^\top R_{2t}
\Big|
&=
n^{1/2}
\Big|
\bbP\bigl[
\Delta\widehat h_t(H_t)\,
\Delta\widehat\pi_t(U_t)\,
{a}^\top\widetilde X_t
\bigr]
\Big| \\
&\le
n^{1/2}
\Bigl[
\bbP\{\Delta\widehat h_t( H_t)\Delta\widehat\pi_t(U_t)\}^2
\Bigr]^{1/2}
\Bigl[
\bbP\{{a}^\top\widetilde X_t\}^2
\Bigr]^{1/2}.
\end{align*}
By Condition 3,
\[
\bbP\{{a}^\top\widetilde X_t\}^2
=
{a}^\top \Gamma_t {a}
\le
\sigma_{\max}^2\|{a}\|_2^2,
\]
and by Condition 4,
\[
\bbP\{\Delta\widehat h_t( H_t)\Delta\widehat\pi_t(U_t)\}^2
=
o_p(n^{-1}).
\]
Hence \(R_{2t}=o_p(1)\).

\medskip
\noindent\textbf{Step 3:} $R_{3t}=o_p(1)$.

We first note that $\bS_t(\bbeta_{0t}, \wh{  {\eta}}_t) - \bS_t(\bbeta_{0t},   {\eta}_{0t}) = \bQ_{1t} + \bQ_{2t}$, where
\begin{align*}
    \bQ_{1t} = W_{t} \Delta \wh\pi_t(U_{t}) \wt X_{t} \quad\mbox{and}\quad
    \bQ_{2t} = \left\{ (V_t + \Delta \wh\pi_t(U_{t})) \Delta \wh{h}_t(H_{t}) \right\} \wt X_{t}.
\end{align*}

Next, we note that 
\begin{align*}
    \mathrm{Trace} \left\{\mathbb{P}\left(Q_{1t}^{\otimes2} \right) \right\} & = \Big| {1}^{\top} \cdot \bbP \left\{  \bQ_{1t} \otimes \bQ_{1t} \right\} \Big| \\
    & = \Big| \bbP  W_{t}^2 \left\{ \Delta \wh\pi_t(U_{t}) \right\}^2 \cdot  {1}^{\top} ( \wt X_{t} \otimes \wt X_{t}) \Big| \\
     &\le c \cdot \bbP \left\{  \left\{ \Delta \wh\pi_t(U_{t}) \right\}^2 \cdot \Big|  {1}^{\top} (\wt X_{t} \otimes \wt X_{t}) \Big| \right\} \nonumber \\
    & \le c d \kappa_{\max}^2  \cdot \left[ \bbP \left\{ \Delta \wh\pi_t(U_{t}) \right\}^4 \right]^{1/2}. 
\end{align*}

Similarly, note that $V_t \in [-1, 1]$, $\Delta \wh{\pi}_t(U_{t}) \in [-1, 1]$, then
\begin{align*}
    \mathrm{Trace} \left\{\mathbb{P}\left(Q_{2t}^{\otimes2} \right) \right\} & \le  4 d \kappa_{\max}^2 \cdot \left[\bbP \left\{ \Delta \wh{h}_t(H_{t}) \right\}^4 \right]^{1/2}.
\end{align*}

Under Condition 3, both $\mathbb{P}Q_{1t}^{\otimes2} $ and $\mathbb{P}Q_{2t}^{\otimes2}$ are of order $o_p(1)$. Furthermore, because ${\cal F}_{\pi}$ and ${\cal F}_{h_t}$ are both $P$-Donsker, $\bbG_n \bQ_{1t}$ and $\bbG_n \bQ_{2t}$ converge in distribution to zero-mean Gaussian distributions with covariance matrices $\bbP \bQ_{1t}^{\otimes 2} $ and $\bbP \bQ_{2t}^{\otimes 2} $, respectively.
As a result, $ {R}_3 = \bbG_n \bQ_{1t} + \bbG_n \bQ_{2t} = O_p(1) \cdot o_p(1) = o_p(1)$.

\medskip
\noindent\textbf{Step 4:} Asymptotic linear representation.

Combining \eqref{eq:proof_basic_expansion} with Steps 1--3 yields
\[
n^{1/2}(\widehat\bbeta_t-\bbeta_{0t})
=
J_{0t}^{-1}\bbG_n S_t(\bbeta_{0t},\eta_{0t})
+
o_p(1).
\]
Stacking these expansions over \(t=1,\dots,T\), we obtain
\[
n^{1/2}(\widehat\bbeta-\bbeta_0)
=
\Gamma_0^{-1}
\begin{pmatrix}
\bbG_n S_1(\bbeta_{01},\eta_{01})\\
\vdots\\
\bbG_n S_T(\bbeta_{0T},\eta_{0T})
\end{pmatrix}
+
o_p(1),
\]
where
\[
\Gamma_0=\mathrm{diag}(J_{01},\dots,J_{0T}).
\]

Since \(\mathcal T\) is finite, the multivariate central limit theorem implies
\[
\begin{pmatrix}
\bbG_n S_1(\bbeta_{01},\eta_{01})\\
\vdots\\
\bbG_n S_T(\bbeta_{0T},\eta_{0T})
\end{pmatrix}
\overset{d}{\longrightarrow}
N(0,\Psi_0),
\]
where
\[
\Psi_0=(\Omega_{0,tt'})_{t,t'\in\mathcal T},
\qquad
\Omega_{0,tt'}
=
E\!\left[
S_t(\bbeta_{0t},\eta_{0t})
S_{t'}(\bbeta_{0t'},\eta_{0t'})^\top
\right].
\]
Therefore,
\[
n^{1/2}(\widehat\bbeta-\bbeta_0)
\overset{d}{\longrightarrow}
N\!\left(
0,\,
\Gamma_0^{-1}\Psi_0\Gamma_0^{-1}
\right).
\]
This completes the proof.
\hfill $\square$

\vspace{2em}
\noindent\textbf{Proof of Theorem 2}: The proof follows the same arguments as the above proof to obtain
\begin{align*}
    n^{1/2} \left( \widehat \bbeta_t - \bbeta_{0t} \right) & = n^{1/2} \left\{ \widehat{ {J}}_{t}^{-1} \Psi_t^{b} \left( \widehat{  {\eta}}_t \right) - \bbeta_{0t}  \right\} \\
    & = n^{1/2} \left( {J}_{0t}^{-1} +  {R}_{1t} \right) 
    \bbP_n S_t(\bbeta_{0t},\widehat\eta_t) \\
    & = \left( {J}_{0t}^{-1} +  {R}_{1t} \right) 
    \Big\{ \bbG_n \bS_t(\bbeta_{0t}, \wh{  {\eta}}_{t})  +  {R}_{2t} \Big\} \\
    & = \left( {J}_{0t}^{-1} +  {R}_{1t} \right) 
    \Big\{ \bbG_n \bS_t(\bbeta_{0t},   {\eta}_{0t})  +  {R}_{2t} +  {R}_{3t} \Big\},
\end{align*}
where $ {R}_{1t}= \widehat{ {J}}_{t}^{-1}  - {J}_{0t}^{-1}$, $ {R}_{2t} = n^{1/2} \bbP {\bS}_t(\bbeta_{0t}, \wh{  {\eta}}_t)$, $ {R}_{3t} = \bbG_n \left\{ \bS_t(\bbeta_{0t}, \wh{  {\eta}}_t) - \bS_t(\bbeta_{0t},   {\eta}_{0t}) \right\}$. 
Note that $ {R}_{1t}$ and $ {R}_{2t}$ are $o_p(1)$ uniformly in $t$ for $t=1,...,T$ because of the uniform convergence of $\widehat\pi_t$ and $\widehat \delta_t$ in Condition 3 and the Donsker property of $ {S}_t(\bbeta_{0t},  \wh{  {\eta}}_t)$. Since $A_t, Y_t$, $U_t$, and $\widetilde \bX_t$ are the realization of some random processes that have bounded total variations and Condition 4 holds, Both $Q_{1t}$ and $Q_{2t}$ as a function of $t$ belong to some Donsker class so $ {R}_{3t}=\sqrt n O_p( (\sup_t\Vert \wh{  {\eta}}_{t}-  {\eta}_{0t}\Vert^4 )^{1/2} )$. 
Therefore, we obtain
$$\sup_{t} \left| n^{1/2} \left( \widehat \bbeta_t - \bbeta_{0t} \right) 
     -  {J}_{0t}^{-1}
    \bbG_n \bS_t(\bbeta_{0t},   {\eta}_{0t}) \right| = o_p(1).$$
Equivalently, it gives
\vspace{-1em}
$$\sup_{s\in \{1/T, 2/T ,...\}} \left| n^{1/2} \left( \wh\bxi(s)-\bxi_0(s) \right) 
     - 
    \bbG_n \widetilde\bGamma(s) \right|= o_p(1).$$
Therefore, by the definition of $\wh\bxi$ and the continuous differentiability of $\bxi_0$, we have
\begin{eqnarray*}
    & &
\sup_{s\in[0,1]} \left|n^{1/2} \left( \wh\bxi(s)-\bxi_0(s) \right) 
     - 
    \bbG_n \widetilde\bGamma(s) \right|\\
    &\le &
    \sup_{s\in \{1/T, 2/T ,...\}} \left|n^{1/2} \left( \wh\bxi(s)-\bxi_0(s) \right) 
     - 
    \bbG_n \widetilde\bGamma(s) \right|\\
    & &+\sup_{t=1,...,T}\sup_{s\in [(t-1)/T, t/T]}\left\{\sqrt n \left|\bxi_0(s)-\bxi_0(t/T) \right|
    + \left|\bbG_n \left(\widetilde\bGamma(s)-\widetilde\bGamma(t/T)\right) \right|\right\}\\
    &\le &    
   \sup_{t=1,...,T}\sup_{s\in [(t-1)/T, t/T]} \left|\bbG_n \left(\widetilde\bGamma(s)-\widetilde\bGamma(t/T)\right) \right|
   +o_p(1)+O(\sqrt n/T).
   \end{eqnarray*}
   Theorem 2 follows from the fact that $\sqrt n/T\rightarrow 0$ and the pointwise convergence of the covariance function for $\widetilde\bGamma(s)$ to the continuous function $\widetilde\bSigma_0$.
   $\square$

\section*{Appendix B}

Appendix B proposes a locally pooled estimator that improves the efficiency of nuisance-function estimation by borrowing information from nearby previous time points under a mild temporal smoothness assumption.

To motivate the construction, recall that at time $t$, untreated observations with $A_{it}=0$ provide direct information on $\mu_{0t}(U_{it})$ through the residualized outcome 
\begin{align*}
    \widehat R_{it}^{(0)}
=
Y_{it}
-
\widehat g(Z_i)
-
\sum_{k=1}^{t-1} A_{ik}\gamma^{t-k}\widehat\beta_{k}^{\top}\widetilde X_{ik}.
\end{align*}
For earlier time points $k<t$, the treatment effect parameter $\beta_{0k}$ has already been estimated in the sequential procedure. Hence, after subtracting the estimated treatment contribution, both treated and untreated observations at time $k$ can be incorporated to inform the nuisance function at nearby time points. This motivates estimating a time-indexed regression function $\mu(u,k)$ by pooling observations over a local neighborhood of times preceding $t$. In practice, $\mu(u,k)$ may be modeled flexibly using, for example, a recurrent neural network to capture smooth temporal evolution.

Specifically, for some lower index $t_0<t$, consider the estimator obtained by minimizing
\begin{align}
\frac{1}{n}\sum_{i=1}^n
\Biggl[
I(A_{it}=0)\bigl\{\widehat R_{it}^{(0)}-\mu(U_{it},t)\bigr\}^2
+
\sum_{k=t_0}^{t-1}\lambda_{t-k}
\bigl\{
\widehat R_{ik}^{(0)}-A_{ik}\widehat\beta_k^{\top}\widetilde X_{ik}-\mu(U_{ik},k)
\bigr\}^2
\Biggr],
\label{eq:local_mu_estimator}
\end{align}
where the weights $\lambda_{t-k}$ satisfy
\[
1 \ge \lambda_1 \ge \lambda_2 \ge \cdots \ge 0.
\]
These weights control the amount of borrowing across time, with observations closer to time \(t\) receiving larger weight. 
We then take $\widehat\mu_t(u)=\widehat\mu(u,t)$ as the resulting estimator of $\mu_{0t}(u)$.

The benefit of \eqref{eq:local_mu_estimator} is a reduction in variance. By borrowing information from nearby time points, the effective sample size used to estimate $\mu_{0t}$ increases, which can substantially stabilize the nuisance estimation when the within-time-point sample size is small or when treatment assignment is highly imbalanced. This can in turn improve the finite-sample performance of the treatment effect estimator $\widehat\beta_t$, since the latter depends on the quality of the nuisance estimate.
A rigorous theoretical analysis of the efficiency gain afforded by this local pooling strategy is left for future work.

\section*{Appendix C}

The Appendix C contains additional Figures for the simulation studies.

\begin{figure}[ht]
    \centering
    \includegraphics[width=1.05\textwidth]{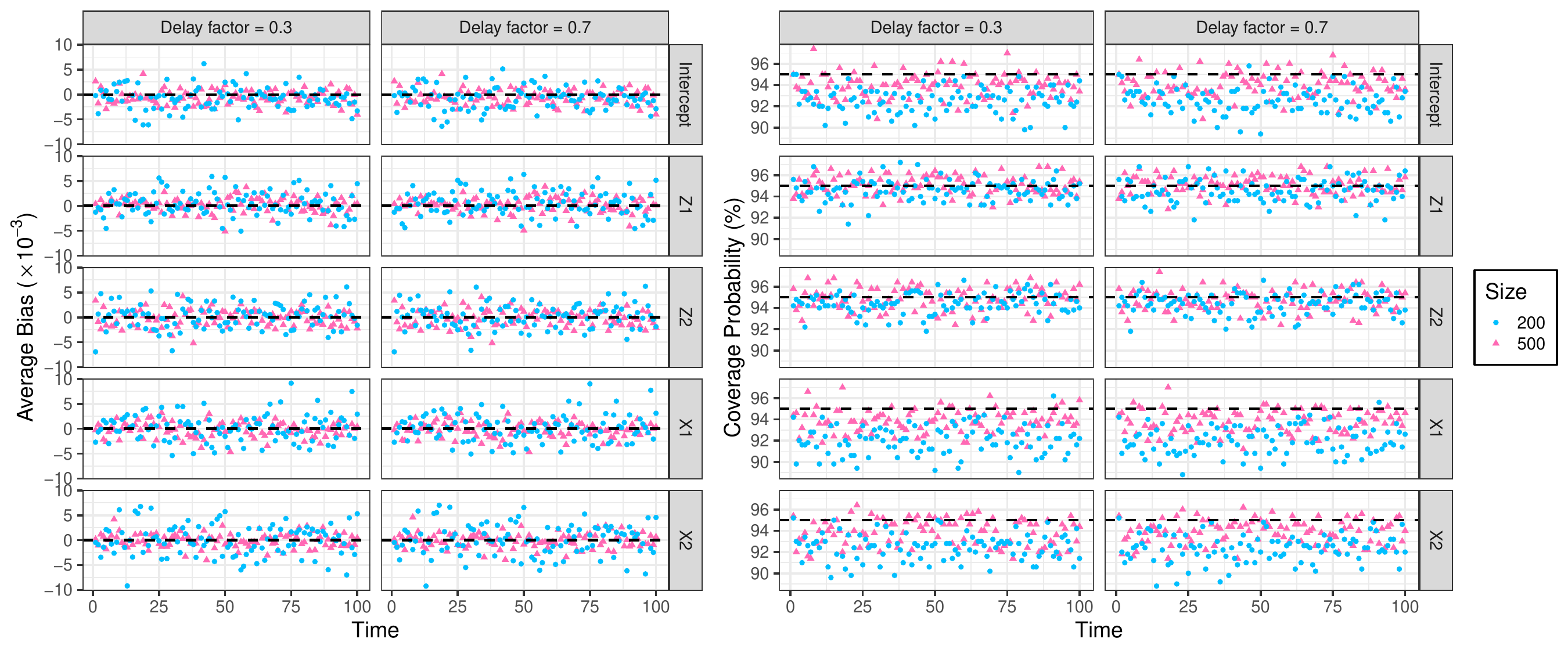}
    \caption{Performance of the proposed method (with known carryover discounting factor) over time across 500 simulations for Case I. The left panel shows the average bias, and the right panel displays the coverage probabilities.}
    \label{img:simu_case1_b}
\end{figure}

\end{document}